\documentclass[lettersize,journal]{IEEEtran}
\usepackage{amsmath,amsfonts}
\usepackage{algorithmic}
\usepackage{algorithm}
\usepackage{array}
\usepackage[caption=false,font=normalsize,labelfont=sf,textfont=sf]{subfig}
\usepackage{textcomp}
\usepackage{stfloats}
\usepackage{hyperref}
\usepackage{url}
\usepackage{verbatim}
\usepackage{graphicx}
\usepackage{cite}
\usepackage{amsmath,amssymb,amsfonts}
\usepackage{bm}
\usepackage[table,xcdraw]{xcolor}
\begin{document}

\title{A Dimension-Decomposed Learning Framework for Online Disturbance Identification in Quadrotor SE(3) Control}

\author{Tianhua Gao
\thanks{This manuscript presents preliminary follow-up progress to the previous works \cite{2025 Robustness Enhancement for Multi-Quadrotor Centralized Transportation System via Online Tuning and Learning}, \cite{2025 Dimension-Decomposed Learning for Quadrotor Geometric Attitude Control with Exponential Convergence on SO(3)} and is not intended for publication. Contact : $^{1}$gao.tianhua@ieee.org.}}

\maketitle

\begin{abstract}
Quadrotor stability under complex dynamic disturbances and model uncertainties poses significant challenges. One of them remains the underfitting problem in high-dimensional features, which limits the identification capability of current learning-based methods. To address this, we introduce a new perspective: Dimension-Decomposed Learning (DiD-L), from which we develop the Sliced Adaptive-Neuro Mapping (SANM) approach for geometric control. Specifically, the high-dimensional mapping for identification is axially ``sliced" into multiple low-dimensional submappings (``slices"). In this way, the complex high-dimensional problem is decomposed into a set of simple low-dimensional tasks addressed by shallow neural networks and adaptive laws. These neural networks and adaptive laws are updated online via Lyapunov-based adaptation without any pre-training or persistent excitation (PE) condition. To enhance the interpretability of the proposed approach, we prove that the full-state closed-loop system exhibits arbitrarily close to exponential stability despite multi-dimensional time-varying disturbances and model uncertainties. This result is novel as it demonstrates exponential convergence without requiring pre-training for unknown disturbances and specific knowledge of the model.  
\end{abstract}


\section{Introduction}

\IEEEPARstart{Q}{uadrotor} stability remains a critical issue under complex disturbances and model uncertainties.
Current research can be broadly categorized into two primary orientations: conventional adaptive control (e.g., \cite{2018 Active Disturbance Rejection Attitude Control for a Dual Closed-Loop Quadrotor Under Gust Wind}-\cite{2025 L1Adaptive Augmentation of Geometric Control for Agile Quadrotors With Performance Guarantees}) and learning-based control (e.g., \cite{2020 Learning-Based Robust Tracking Control of Quadrotor With Time-Varying and Coupling Uncertainties}-\cite{2024 Active Learning of Discrete-Time Dynamics for Uncertainty-Aware Model Predictive Control}). Each category has its own advantages and limitations that need to be addressed.

Conventional adaptive methods generally provide better interpretability since they typically rely on explicit system structure or disturbance modeling. However, their performance may be limited when dealing with highly nonlinear disturbances, such as turbulent wind fields.  For example, in \cite{2022 Performance Precision and Payloads Adaptive Nonlinear MPC for Quadrotors, 2023 Active Wind Rejection Control for a Quadrotor UAV Against Unknown Winds},  wind effects are handled using linear drag coefficients and compensated by adaptive laws. In \cite{2018 Active Disturbance Rejection Attitude Control for a Dual Closed-Loop Quadrotor Under Gust Wind}, \cite{2024 Experimental Validation of a Robust Prescribed Performance Nonlinear Controller for an Unmanned Aerial Vehicle With Unknown Mass, 2024 Quadrotor Fault-Tolerant Control at High Speed: A Model-Based Extended State Observer for Mismatched Disturbance Rejection Approach}, the authors employ the Extended State Observer (ESO) to compensate for disturbances without explicit disturbance modeling, but it remains dependent on assumed structural properties. 

In contrast, learning-based methods leverage neural networks to  better approximate complex nonlinear features. These methods have been extensively validated through experiments and demonstrate fast convergence properties \cite{2020 Learning-Based Robust Tracking Control of Quadrotor With Time-Varying and Coupling Uncertainties}-\cite{2024 Neural Moving Horizon Estimation for Robust Flight Control}. However, the approximation error of neural networks remains a significant concern, as these methods typically adopt shallow neural networks (SNN), which tend to underfit high-dimensional features during online process. To tackle this, recent studies on data-driven control leverage the representation power of deep neural networks (DNN) for precise offline identification of disturbance and uncertainty features \cite{2022 Neural-Fly enables rapid learning for agile flight in strong winds}-\cite{2024 Meta-Learning Augmented MPC for Disturbance-Aware Motion Planning and Control of Quadrotors}. These methods exhibit great potential, but issues remain in the weak interpretability of offline training process and generalization capability to unseen environments. Therefore, we attempt to develop a direct enhancement of learning-based methods without relying on data-driven identification.

In this paper, we present a new branch in learning-based control: Dimension-Decomposed Learning (DiD-L). The key idea is to decompose high-dimensional disturbances and uncertainties into multiple lower-dimensional features, which specifies and simplifies the task of each SNN. Our contributions in this work are summarized as follows:

\textbf{\textit{(1)}} Proposed the first DiD-L instance: Sliced Adaptive-Neuro Mapping (SANM) with the following advantages:
\begin{itemize}
\item \textbf{\textit{Full-state Compensation}}-Some existing studies (e.g., \cite{2023 Active Wind Rejection Control for a Quadrotor UAV Against Unknown Winds, 2024 Experimental Validation of a Robust Prescribed Performance Nonlinear Controller for an Unmanned Aerial Vehicle With Unknown Mass, 2024 Predictor-Based Neural Attitude Control of A Quadrotor With Disturbances, 2022 Neural-Fly enables rapid learning for agile flight in strong winds, 2024 Active Learning of Discrete-Time Dynamics for Uncertainty-Aware Model Predictive Control }) only addressed force disturbance or moment disturbance, this work presents full-state compensation for multi-dimensional disturbances (both force and moment) and model uncertainties.
\item $\mathbf{SE}(3)$ \textbf{Compatibility}-SANM can be deployed onto existing geometric control on $\mathbf{SE}(3)$\cite{2010 Geometric tracking control of a quadrotor UAV on SE(3)}, which does not rely on small-angle assumptions \cite{2023 Active Wind Rejection Control for a Quadrotor UAV Against Unknown Winds} or linearized models.
\item \textbf{\textit{Highly Customizable}}-The adaptive law and SNN on each slice can be individually customized based on the dynamic characteristics of different dimensions. Moreover, while this work proposes a 12-slice SANM, the number of slices can be flexibly adjusted according to disturbance rejection requirements during actual deployment.
\item \textbf{\textit{Efficient Representation}}-After dimension decomposition, only 5 neurons in a single layer achieve an effective approximation to unseen disturbance in each dimension. 
\item \textbf{\textit{Rapid Response}}-SANM learns disturbance features at the acceleration-level, thereby achieving a transient response.
\item \textbf{\textit{Strong Generalization}}-SNNs are updated online via Lyapunov-based adaptation, ensuring bounded  weight estimation in unseen environments without persistent excitation (PE) condition.
\item \textbf{\textit{Strong Interpretability}}-A rigorous Lyapunov analysis that explicitly considers neural network approximation errors supports the interpretability of SANM.
\item \textbf{\textit{Exponential Convergence}}- All state errors exponentially converge to an arbitrarily small ball.
\end{itemize}

\textbf{\textit{(2)}} Proved the Near-Exponential Stability (NES) of the proposed control system, a new concept defined in \textbf{\textit{Definition 1}}, which is arbitrarily close to exponential stability. To the best of our knowledge, this result is novel in quadrotor learning-based control against disturbances and uncertainties.

\textbf{\textit{(3)}} Demonstrated the feasibility and advantages of SANM, through real-time simulation experiments performed in Gazebo, a high-fidelity physics simulator.

This paper is organized as follows. Section \uppercase\expandafter{\romannumeral 2} describes the problem formulation. Section \uppercase\expandafter{\romannumeral 3} introduces the design of SANM and controller. Section \uppercase\expandafter{\romannumeral 4} presents results of physics simulation experiments. Finally, Section \uppercase\expandafter{\romannumeral 5} concludes the paper and discusses future work. The stability proof is supplemented in Appendix.

\section{Problem Formulation}
\subsection{Quadrotor Dynamics with Augmented Disturbance }
\begin{figure}[!t]
      \centering
      \includegraphics[scale=0.18]{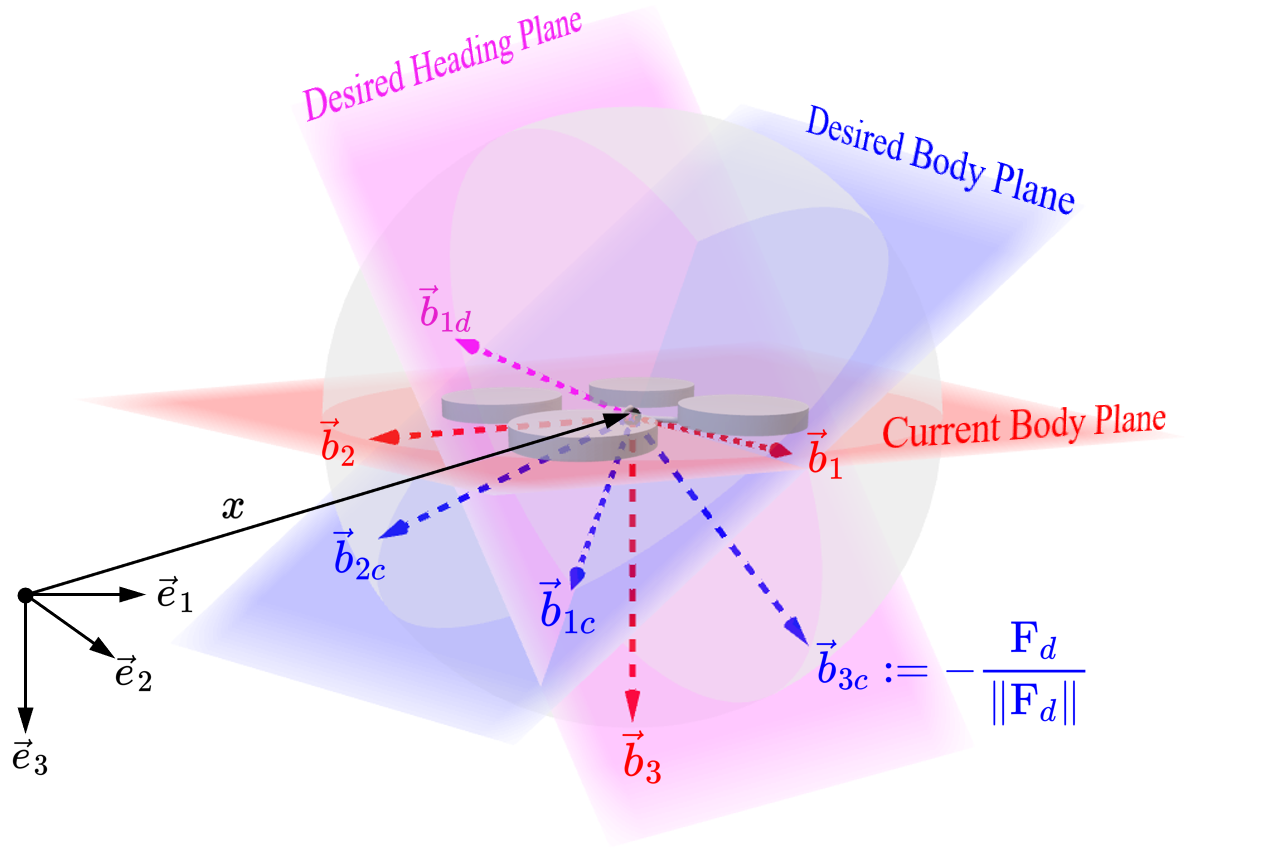}
      \caption{\footnotesize Quadrotor modeling and geometric control on $\mathbf{SE}(3)$. The vectors $\bm{\vec{b}}_{1\bm{d}}$, $\bm{\vec{b}}_{1\bm{c}}$ and $\bm{\vec{b}}_{3\bm{c}}$ are coplanar and form the desired heading plane, while the vectors $\bm{\vec{b}}_{1\bm{c}}$ and  $\bm{\vec{b}}_{2\bm{c}}$ are coplanar and form the desired body plane. }
      \label{Gao1}
   \end{figure}

This section introduces the dynamics of the quadrotor augmented by disturbances. A North-East-Down (NED) inertia frame $\mathcal{I}:=\{\bm{\vec{e}}_j\}_{1\leq j \leq 3}$ and an NED quadrotor body-fixed frames  $\mathcal{B}:=\{\bm{\vec{b}}_{j}\}_{1\leq j \leq 3}$ are defined as shown in Fig. \ref{Gao1}. The quadrotor model is considered as a rigid body with its center of mass located at the geometric center of the structure, denoted as $\bm{x}\in\mathbb{R}^3$. The orientation of the quadrotor is described by a rotation matrix $\bm{R}\in \mathbf{SO}(3) = \{\bm{R}\in\mathbb{R}^{3\times3}\mid\bm{R}^{\top}\bm{R} = \bm{I}^{3\times3}, \mathrm{det}(R) = 1\}$, which represents the rotation of $\mathcal{B}$ relative to $\mathcal{I}$. 
For disturbance modeling, we consider two scenarios.

\textbf{\textit{Scenario 1: ($\bm{J}$ is known)}} If the inertia tensor $\bm{J}\!\in\!\mathbb{R}^{3\times3}$ is known, we augment the standard quadrotor dynamics with unknown time-varying dynamics  terms of  translational and rotational disturbance $\bm{\phi}_{\bm{\textit{x}}}$, $\bm{\phi}_{\bm{\textit{R}}}\in \mathbb{R}^3$ at the acceleration-level:
\begin{equation}
{
\footnotesize
\left\{
\begin{aligned}
&\bm{\dot{x}}=\bm{v},\\
&\bm{\dot{v}}=\frac{1}{m}\bm{\mathrm{F}_d}+g\bm{\vec{e}}_3+\bm{\phi}_{\bm{\textit{x}}},\\
&\bm{\dot{R}}= \bm{R}[\bm{\Omega}]_{\times},\\
&\bm{\dot{\Omega}} = \bm{J}^{\scalebox{0.6}{$-1$}}\left(\bm{\mathrm{M}_d}-[\bm{\Omega}]_{\times}\bm{J}\bm{\Omega}\right)+\bm{\phi}_{\bm{\textit{R}}},\\
\end{aligned}
\right.
\!\!\!\!\!\!{\to}
\left\{
\begin{aligned}
&\bm{\dot{x}}=\bm{v},\\
&\bm{\dot{v}}=-\frac{1}{m}f\bm{R}\bm{\vec{e}}_3+g\bm{\vec{e}}_3+\bm{\phi}_{\bm{\textit{x}}},\\
&\bm{\dot{R}}= \bm{R}[\bm{\Omega}]_{\times},\\
&\bm{\dot{\Omega}} = \bm{J}^{\scalebox{0.6}{$-1$}}\left(\bm{\mathrm{M}}-[\bm{\Omega}]_{\times}\bm{J}\bm{\Omega}\right)+\bm{\phi}_{\bm{\textit{R}}},\\
\end{aligned}
\right.
\label{Dynamics with Augmented Disturbance}}
\end{equation}
where the equations on the left are the ideal dynamics with desired resultant control force $\bm{\mathrm{F}_d}\in\mathbb{R}^3$ and moment $\bm{\mathrm{M}_d}\in\mathbb{R}^3$. Through the control flow in Section \ref{sec::Control Formulation Overview}, desired control wrench $\{\bm{\mathrm{F}_d},\bm{\mathrm{M}_d}\}$  is transformed into motor commands, which generate the actual resultant thrust $f\in\mathbb{R}$ and moment $\bm{\mathrm{M}}\in\mathbb{R}^3$ included within the real dynamics on the right. The linear velocity in the inertial frame is denoted by $\bm{v}\in\mathbb{R}^3$, and $\bm{\Omega}\in\mathbb{R}^3$ is the anugular velocity in the body-fixed frame. Symbol $[\,\bullet\,]_{\times}\!:\!\mathbb{R}^3\!\to\!\mathfrak{so}(3)$ represents the skew-symmetric map defined by the condition that $[\mathfrak{a}]_{\times}\mathfrak{b}=\mathfrak{a}\times\mathfrak{b}, \forall\mathfrak{a},\mathfrak{b}\in\mathbb{R}^3$. The gravitational acceleration $g\in\mathbb{R}$ is a constant scalar, while the mass  $m\!\in\!\mathbb{R}$ is considered to be unknown variable.

\textbf{\textit{Scenario 2: ($\bm{J}$ is unknown)}} If the inertia tensor is unknown, the term $\bm{J}^{-1}[\bm{\Omega}]_{\times}\bm{J}\bm{\Omega}$ in Eq.~\eqref{Dynamics with Augmented Disturbance} cannot be compensated for in the attitude control introduced later in Eq.~\eqref{Md}. However, since this term also represents an unknown time-varying dynamic component, we can treat it as an internal disturbance and incorporate it into the  $\bm{\phi}_{\bm{\textit{R}}}$ term:
\begin{equation}
    \bm{\dot{\Omega}} = \bm{J}^{\scalebox{0.6}{$-1$}}\bm{\mathrm{M}}+\bm{\phi}_{\bm{\textit{R}}}(\bm{J}, \bm{\Omega}),
    \label{Dynamics with Augmented Disturbance2}
\end{equation}
where $\bm{\phi}_{\bm{\textit{R}}}(\bm{J}, \bm{\Omega})$ represents the total unknown rotational disturbance, including both internal and external disturbances. 

\begin{table}[!t]
  \centering
  \caption{List of Notations: Maps, Subscripts and Superscripts}
  \begin{tabular}{>{\columncolor{gray!20}}c|l}
    \hline
    \hline
    $[\,\bullet\,]_{\times}$ & Skew-symmetric map: $\mathbb{R}^3\to\mathfrak{so}(3)$\\
   $\bullet^{[\cdot]}$ &  Element extraction map: $ (\mathbb{R}^3\cup\mathbb{R}^{3\times 3})\times\mathbb{N}\to\mathbb{R}$\\
   $\bullet^\vee$& Vee map: $\mathfrak{so}(3)\to\mathbb{R}^3$\\
   $\bullet_{\bm{d}}$& Desired value (given)\\
   $\bullet_{\bm{c}}$& Desired value (computed)\\
   $\bullet^{\text{rec}}$ & Reconstructed feature vector\\
   $\bar{\bullet}$ & Estimation value\\
   $\widetilde{\bullet}$ & Estimation error value\\
    $\lambda_{\min}(\bullet)$ &Minimum eigenvalue of a matrix \\
  $\lambda_{\max}(\bullet)$ &Maximum eigenvalue of a matrix\\
    \hline
    \hline
  \end{tabular}
  \label{table1}
\end{table}
\subsection{Control Problem Formulation on $\mathbf{SE}(3)$.}
The position, velocity, attitude and angular velocity tracking errors of quadrotor system $\bm{e}_{\bm{x}}$, $\bm{e}_{\bm{v}}$, $\bm{e}_{\bm{R}}$, $\bm{e}_{\bm{\Omega}}\in\mathbb{R}^3$ are defined and summarized as follows:
\begin{equation}
\begin{aligned}
\bm{e}_{\bm{x}}:=&\bm{x}-\bm{x}_{{\bm{d}}}, \, \,\,\bm{e}_{\bm{v}}:=\bm{v}-\bm{\dot{x}}_{{\bm{d}}}=\bm{\dot{e}}_{\bm{x}}, \\
     \bm{e}_{\bm{R}}:=&\frac{1}{2}(\bm{R}_{{\bm{c}}}^{\top}\bm{R}-\bm{R}^{\top}\bm{R}_{{\bm{c}}})^\vee , \bm{e}_{\bm{\Omega}}:=\bm{\Omega}-\bm{R}^{\top}\bm{R}_{{\bm{c}}}\bm{\Omega}_{{\bm{c}}},
    \label{errors}
\end{aligned}
\end{equation}
where vee map $\bullet^\vee: \mathfrak{so}(3)\to\mathbb{R}^3$ denotes the inverse of skew-symmetric map $[\,\bullet\,]_{\times}$. The $\bm{x}_{{\bm{d}}}(t)\in\mathbb{R}^3$ is the desired position and $\bm{R}_{{\bm{c}}}\in \mathbf{SO}(3)$ is the desired attitude computed by giving the desired heading direction $\bm{\vec{b}}_{1\bm{d}}(t)\in\mathbf{S}^2$: 
\begin{equation}
    \bm{R}_{{\bm{c}}}:=[\bm{\vec{b}}_{1\bm{c}},\bm{\vec{b}}_{2\bm{c}},\bm{\vec{b}}_{3\bm{c}}],
    \label{R_c}
\end{equation}
with $\bm{\vec{b}}_{1\bm{c}}:=\bm{\vec{b}}_{2\bm{c}}\times\bm{\vec{b}}_{3\bm{c}}$, $\bm{\vec{b}}_{2\bm{c}}:=(\bm{\vec{b}}_{3\bm{c}}\times\bm{\vec{b}}_{1\bm{d}})/(\|\bm{\vec{b}}_{3\bm{c}}\times\bm{\vec{b}}_{1\bm{d}}\|)$ and $\bm{\vec{b}}_{3\bm{c}}:= -\bm{\mathrm{F}_d}/\|\bm{\mathrm{F}_d}\|$, as shown in Fig. \ref{Gao1}. The desired angular velocity can then be further computed through differentiation:
\begin{equation}
    \bm{\Omega}_{{\bm{c}}}:=(\bm{R}^{\top}_{{\bm{c}}}\dot{\bm{R}_{{\bm{c}}}})^{\vee}.
    \label{Omegac}
\end{equation}

The above state error formulation follows the standard geometric tracking control structure on $\mathbf{SE}(3)$ for quadrotor systems. Our objective is to design the 6-dimensional desired control wrench $\{\bm{\mathrm{F}_d},\bm{\mathrm{M}_d}\}$, given a bounded desired trajectory
$\bm{x}_{{\bm{d}}}(t)$ and a desired heading direction $\bm{\vec{b}}_{1\bm{d}}(t)$. The aim is to achieve the exponential convergence of all state errors $\bm{e}_{\bm{x}}$, $\bm{e}_{\bm{v}}$, $\bm{e}_{\bm{R}}$, $\bm{e}_{\bm{\Omega}}$ with the presence of unknown but bounded model parameters $\{m,\bm{J}\}$  and bounded 6-dimensional disturbance dynamics $\{\bm{\phi}_{\bm{\textit{x}}}, \bm{\phi}_{\bm{\textit{R}}}\}$  acting on dynamics in $\mathbf{SE}(3)$. 

\section{ Sliced Adaptive-Neuro Mapping for Geometric Control on $\mathbf{SE}(3)$}
\label{sec::High-level Control Design}

\subsection{Quadrotor Geometric Control on $\mathbf{SE}(3)$}
\label{sec::Control Formulation Overview}

\begin{figure}[!t] 
 \centering
 \includegraphics[scale=0.125]{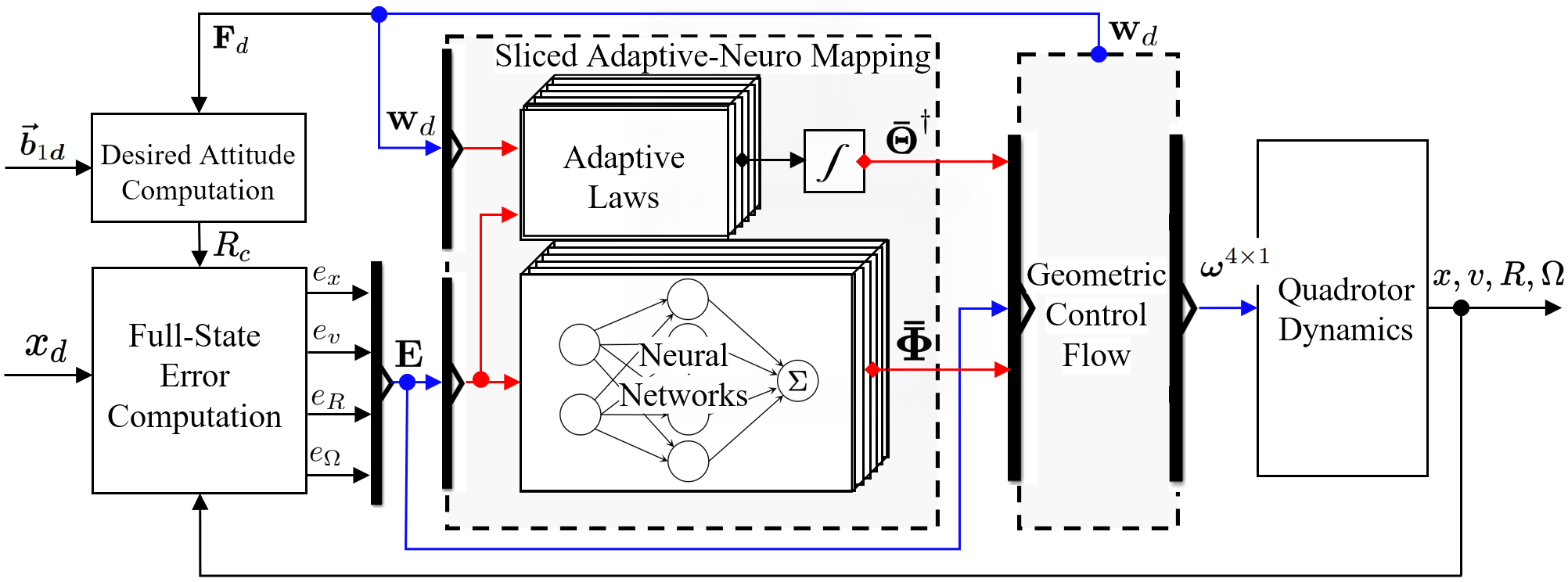}
 \caption{\footnotesize The  structure of  SANM-geometric control  strategy. } 
      \label{Gao3}
\end{figure}

Our control system adopts a multi-level geometric control flow (see Fig. \ref{Gao3}), involving hierarchical control signal transformations from high-level commands to low-level control inputs:
\begin{align}
\to\bm{\mathrm{w}_d}^{6\times1}\to{\renewcommand{\arraystretch}{1}\begin{pmatrix}f_d \\ \bm{\mathrm{M}_d}\\\end{pmatrix}}^{4\times1}\!\!\!\!\!\!\to \bm{T}_{\bm{d}}^{4\times1}\to\bm{\omega}^{4\times1}\!\!\to {\renewcommand{\arraystretch}{1}\begin{pmatrix}f \\ \bm{\mathrm{M}}\\\end{pmatrix}}^{4\times1}\!\!\!\!\!\!\!\!\to,
\end{align}
where $\bm{\mathrm{w}_d}=(\bm{\mathrm{F}_d}^{\top},\bm{\mathrm{M}_d}^{\top})^{\top}\in\mathbb{R}^{6}$ is the desired control wrench vector, which will be designed later as high-level commands in Section \ref{Wrench Controller Design}. $f_d\in\mathbb{R}$ is the desired total thrust projected from the desired resultant thrust $\bm{\mathrm{F}_d}$ onto the body-fixed frame $\bm{\vec{b}}_{3}$ axis: 
\begin{equation}
    f_d:=-\bm{\mathrm{F}_d} \cdot \bm{R}\bm{\vec{e}}_3,
\end{equation}
and the desired thrusts for each rotor $\bm{T}_{\bm{d}}\in\mathbb{R}^{4}=[T_{d1},T_{d2},T_{d3},T_{d4}]^{\top}$ are then computed by following allocation mapping for the X-configuration: 
\begin{equation}
    \begin{aligned}
\bm{T}^{4\times1}_{\bm{d}}:=\frac{1}{4}{\renewcommand{\arraystretch}{1.2}\begin{bmatrix}1 & \frac{\sqrt{2}}{d}& \frac{\sqrt{2}}{d} & \frac{c'_M}{c'_T} \\ 1 & \scalebox{0.6}{$-$}\frac{\sqrt{2}}{d} & \frac{\sqrt{2}}{d} & \scalebox{0.6}{$-$}\frac{c'_M}{c'_T}\\1 & \scalebox{0.6}{$-$}\frac{\sqrt{2}}{d} & \scalebox{0.6}{$-$}\frac{\sqrt{2}}{d} & \frac{c'_M}{c'_T} \\1 & \frac{\sqrt{2}}{d} &\scalebox{0.6}{$-$}\frac{\sqrt{2}}{d}& \scalebox{0.6}{$-$}\frac{c'_M}{c'_T} \end{bmatrix}}{\renewcommand{\arraystretch}{1.2}\begin{pmatrix}f_d \\ \bm{\mathrm{M}_d}\\\end{pmatrix}}^{4\times1}\!\!\!\!\!\!\!\!,
\end{aligned}
\end{equation}
where $d\in\mathbb{R}$ represents the distance between the center of the body-fixed frame and rotors. $c'_M\in\mathbb{R}$ and $c'_T\in\mathbb{R}$
are constant thrust and moment reference coefficients. The rotor speeds for each motor $\bm{\omega}\in\mathbb{R}^4_{+}=[\omega_1,\omega_2,\omega_3,\omega_4]^{\top}$ are then derived as follows:
\begin{equation}
   \omega_i :=\sqrt{\frac{T_{di}}{c'_T}},
\end{equation}
where $\omega_i$ and $T_{di}$ are the rotor speed and desired thrust of $i^{th}$ motor, respectively. Eventually, the actual resultant thrust $f\in\mathbb{R}$ and moment $\bm{\mathrm{M}}\in\mathbb{R}^3$ generated by four rotors can be expressed through the following mapping:
\begin{align}
    {\renewcommand{\arraystretch}{1}\begin{pmatrix}f \\ \bm{\mathrm{M}}\\\end{pmatrix}}^{4\times1}\!\!\!\!\!=
    \setlength{\arraycolsep}{2pt}{{\renewcommand{\arraystretch}{1.2}\begin{bmatrix} c_{T} & c_{T}& c_{T}& c_{T} \\ \frac{\sqrt{2}}{2}dc_{T} & \scalebox{0.6}{$-$}\frac{\sqrt{2}}{2}dc_{T}& \scalebox{0.6}{$-$}\frac{\sqrt{2}}{2}dc_{T}& \frac{\sqrt{2}}{2}dc_{T}\\\frac{\sqrt{2}}{2}dc_{T} & \frac{\sqrt{2}}{2}dc_{T}& \scalebox{0.6}{$-$}\frac{\sqrt{2}}{2}dc_{T}& \scalebox{0.6}{$-$}\frac{\sqrt{2}}{2}dc_{T}\\c_{M} & \scalebox{0.6}{$-$}c_{M}& c_{M}& \scalebox{0.6}{$-$}c_{M} \end{bmatrix}}} {\renewcommand{\arraystretch}{1.2}\begin{pmatrix}\omega_{1}^{2}\\ \omega_{2}^{2}\\\omega_{3}^{2}\\\omega_{4}^{2}\\\end{pmatrix}},
\end{align}
where $c_{T}\in\mathbb{R}$ and $c_{M}\in\mathbb{R}$ are constant
thrust and moment physical coefficients of real rotor aerodynamics. The mapping deviations of resultant thrust $\Delta_f\in\mathbb{R}$ and moment $\bm{\Delta_{\mathrm{M}}}\in\mathbb{R}^{3}$ are defined as follows:
\begin{equation}
   \Delta_f\triangleq f-f_d,\,\bm{\Delta_{\mathrm{M}}}\triangleq \bm{\mathrm{M}}-\bm{\mathrm{M}}_d.
   \label{mapping deviations}
\end{equation}
Given that the $\|\bm{\omega}\|$, $\| c_T - c'_T\|$ and $\| c_M - c'_M\|$ are bounded, the $\|\Delta_f\|$  and $\|\bm{\Delta_{\mathrm{M}}}\|$ are also bounded.  And if $ c_T\to c'_T$, $ c_M\to c'_M$, it follows that $\|\Delta_f\|$ and $\|\bm{\Delta_{\mathrm{M}}}\|$ converge to zero.

\subsection{Wrench Controller with Sliced Adaptive-Neuro Mapping}

Consider a nonlinear continuous mapping from the desired control wrench vector $\bm{\mathrm{w}_d}$, unknown model $\bm{\Theta}\triangleq(m,\bm{J})_{\in\mathbb{R}\times\mathbb{R}^{3\times3}}$ and unknown disturbance dynamics $\bm{\Phi}\triangleq\left(\bm{\phi}^{\top}_{\bm{\textit{x}}}, \bm{\phi}^{\top}_{\bm{\textit{R}}}\right)^{\top}\in\mathbb{R}^6$ to the full-state error $\mathbf{E}\triangleq\big{(}\bm{e}^{\top}_{\bm{x}},\bm{e}^{\top}_{\bm{v}},$ $\bm{e}^{\top}_{\bm{R}},\bm{e}^{\top}_{\bm{\Omega}}\big{)}^{\top}\in\mathbb{R}^{12}$, denoted as $\mathbf{E}\triangleq\bm{\mathcal{S}}(\bm{\mathrm{w}_d},\bm{\Theta},\bm{\Phi}):\mathbb{R}^6\times(\mathbb{R}\times\mathbb{R}^{3\times3})\times\mathbb{R}^6\to\mathbb{R}^{12}$. Since $\bm{\mathcal{S}}$ is not bijective, its inverse mapping does not exist. Nevertheless, we assume that its idealized  pseudo-inverse mapping, denoted as $(\bm{\mathrm{w}_d},\bm{\Theta}^{\text{rec}},\bm{\Phi})\triangleq\bm{\mathcal{S}}^\dagger(\mathbf{E}):\bm{\mathcal{C}}\to\mathbb{R}^6\times\mathbb{R}^6\times\mathbb{R}^6$, exists locally with the input $\mathbf{E}$ bounded on a compact set $\bm{\mathcal{C}}\subset\mathbb{R}^{12}$. Here, the reconstructed model $\bm{\Theta}^{\text{rec}}\triangleq\left(m,m,m, \bm{J}^{[1]},  \bm{J}^{[2]},  \bm{J}^{[3]}\right)^{\top}\in\mathbb{R}^6$  is a vector reconstructed from $\bm{\Theta}$. This model assumes axis alignment, where the mass and the moments of inertia are aligned with the principal axes $\{\bm{\vec{e}}_j\}_{1\leq j \leq 3}$ and $\{\bm{\vec{b}}_j\}_{1\leq j \leq 3}$, respectively.

\textbf{\textit{Notation 1:}} Superscript $\bullet^{[\cdot]}$ denotes an element extraction map $\bullet^{[\cdot]}: (\mathbb{R}^3\cup\mathbb{R}^{3\times 3})\times\mathbb{N}\to\mathbb{R}$ which extracts the $\cdot^{th}$ element from either a vector or the main diagonal of a matrix.

  Since $\bm{\mathrm{w}_d}$, $\bm{\Theta}^{\text{rec}}$ and  $\bm{\Phi}$ are mutually independent, three submappings of $\bm{\mathcal{S}}^\dagger(\mathbf{E})$, denoted as $(\bm{\Theta}^{\text{rec}},\bm{\Phi})\triangleq\bm{\mathcal{S}}^\dagger_{\Theta^{\dagger}\Phi}(\mathbf{E}):\bm{\mathcal{C}}\to\mathbb{R}^6\times\mathbb{R}^6$, $\bm{\Theta}^{\text{rec}}\triangleq\bm{\mathcal{S}}^\dagger_{\Theta^{\dagger}}(\mathbf{E}):\bm{\mathcal{C}}\to\mathbb{R}^6$ and $\bm{\Phi}\triangleq\bm{\mathcal{S}}^\dagger_{\Phi}(\mathbf{E}):\bm{\mathcal{C}}\to\mathbb{R}^6$   also exist.  To approximate  $\bm{\mathcal{S}}^\dagger_{\Theta^{\dagger}\Phi}(\mathbf{E})$, an Adaptive-Neuro mapping $\bm{\mathcal{S}}_{AN}(\bm{\mathrm{w}_d},\mathbf{E}):\mathbb{R}^6\times\bm{\mathcal{C}}\to\mathbb{R}^6\times\mathbb{R}^6$  is preliminarily formulated as follows:
\begin{equation}
 (\!\!\!\!\!\!\!\!\!\!\underbrace{\bm{\bar{\Theta}}^{\text{rec}},\bm{\bar{\Phi}}}_{\textbf{Target Feature Space}}\!\!\!\!\!\!\!\!\!\!) \triangleq \bm{\mathcal{S}}_{AN}(\!\!\!\!\!\!\!\!\!\!\underbrace{\bm{\mathrm{w}_d},\mathbf{E}}_{\textbf{Input Feature Space}}\!\!\!\!\!\!\!\!\!\!):\mathbb{R}^6\times\bm{\mathcal{C}}\to\mathbb{R}^6\times\mathbb{R}^6,
\end{equation}
where vector $\bm{\bar{\Theta}}^{\text{rec}}\!\!:=\!\!\left(\bm{\bar{m}}^{[1]},\bm{\bar{m}}^{[2]},\bm{\bar{m}}^{[3]},\bm{\bar{J}}^{[1]},\bm{\bar{J}}^{[2]},\bm{\bar{J}}^{[3]}\right)^{\top}\!\!\in\!\!\mathbb{R}^{6}$ denotes the estimated pseudo model feature and vector $\bm{\bar{\Phi}} \!:=\!\!\left(\bm{\bar{\phi}}_{\bm{\textit{x}}}^{[1]},\bm{\bar{\phi}}_{\bm{\textit{x}}}^{[2]},\bm{\bar{\phi}}_{\bm{\textit{x}}}^{[3]},\bm{\bar{\phi}}_{\bm{\textit{R}}}^{[1]},\bm{\bar{\phi}}_{\bm{\textit{R}}}^{[2]},\bm{\bar{\phi}}_{\bm{\textit{R}}}^{[3]}\right)^{\top}\!\!\in\mathbb{R}^6$ represents the estimated disturbance dynamics feature. Specifically, $\{\bm{\bar{m}}^{[j]}\in\mathbb{R}\}_{1\leq j \leq 3}$ and $\{\bm{\bar{J}}^{[j]}\in\mathbb{R}\}_{1\leq j \leq 3}$ are the estimated values to the mass  along  $\bm{\vec{e}}_{j}$-axis and the moment of inertia along  $\bm{\vec{b}}_{j}$-axis, respectively. The $\{\bm{\bar{\phi}}_{\bm{\textit{x}}}^{[j]}\in\mathbb{R}\}_{1\leq j \leq 3}$ and $\{\bm{\bar{\phi}}_{\bm{\textit{R}}}^{[j]}\in\mathbb{R}\}_{1\leq j \leq 3}$ are the estimated dynamics of translational and rotational disturbance decomposed along  $\bm{\vec{e}}_{j}$-axis and  $\bm{\vec{b}}_{j}$-axis, respectively.

Next, we slice the high-dimensional Adaptive-Neuro mapping into a set of low-dimensional submappings (“slices”), described as follows for $1\leq j \leq 3$:
\begin{equation}
\footnotesize{
\begin{aligned}
&\bigoplus_{j=1}^{3}\Bigg{\{}\Big{(}\underbrace{\bm{\bar{m}}^{[j]},\bm{\bar{J}}^{[j]},\bm{\bar{\phi}}_{\bm{\textit{x}}}^{[j]},\bm{\bar{\phi}}_{\bm{\textit{R}}}^{[j]}}_{\textbf{Sliced Target Feature Space}}\Big{)}\!:=\\
&\! \bm{\mathcal{S}}^{[j]}_{AN}\Big{(}\underbrace{\bm{\mathrm{F}_d}^{[j]},\bm{\mathrm{M}_d}^{[j]},\bm{\mathcal{E}}_{\bm{\textit{x}} j},(\bm{e}_{\bm{R}}^{[j]}, \bm{e}_{\bm{\Omega}}^{[j]})}_{\textbf{Sliced Input Feature Space}}\Big{)}\!:\!\mathbb{R}\!\times\!\!\mathbb{R}\!\times\!\mathbb{R}^{2}\!\times\!\mathbb{R}^{2}\!\!\to\!\mathbb{R}\!\times\!\mathbb{R}\!\times\!\mathbb{R}\!\times\!\mathbb{R}\Bigg{\}}\\
\end{aligned}
}\notag
\end{equation}
where vector $\bm{\mathcal{E}}_{\bm{\textit{x}} j}:= \left(\bm{e}_{\bm{x}}^{[j]}, \bm{e}_{\bm{v}}^{[j]}\right)^{\top}\!\!\!\!\in\mathbb{R}^2$ denotes the translational error vector along $\bm{\vec{e}}_j$-axis. The  structure of this Sliced Adaptive-Neuro Mapping (SANM) is shown in Fig.~\ref{Gao2}. 
The original 12-dimensional target feature is decomposed into twelve sliced 1-dimensional features.
In this way, the original high-dimensional identification problem is transformed into a set of low-dimensional approximations. The design of these twelve slices is detailed in the following subsections.
\subsubsection{\textbf{Wrench Controller Design}}
\label{Wrench Controller Design}
First of all, we consider using the foregoing target features $\{\bm{\bar{m}}^{[j]},\bm{\bar{J}}^{[j]},\bm{\bar{\phi}}_{\bm{\textit{x}}}^{[j]},\bm{\bar{\phi}}_{\bm{\textit{R}}}^{[j]}\}_{1\leq j \leq3}$ from SANM to compensate the disturbances and uncertainties.  To apply these values, 
the desired control wrench  $\{\bm{\mathrm{F}_d},\bm{\mathrm{M}_d}\}$ is axially decomposed into individual components $\{\bm{\mathrm{F}_d}^{[j]},\bm{\mathrm{M}_d}^{[j]}\}_{1\leq j\leq3}$. These components are designed as follows:
\begin{equation}
    \begin{aligned}
\bm{\mathrm{F}_d}^{[j]}:=&\bm{\bar{m}}^{[j]}
\left(-\bm{\mathcal{K}}_{\bm{\textit{x}} j}^{\top}\bm{\mathcal{E}}_{\bm{\textit{x}} j}+\bm{\ddot{x}}_{{\bm{d}}}^{[j]}-g\bm{\delta}_{j3} -\bm{\bar{\phi}}_{\bm{\textit{x}}}^{[j]}\right),\label{Fd}
\end{aligned}
\end{equation}
\begin{equation}
\begin{aligned}
\bm{\mathrm{M}_d}^{[j]}:=&\bm{\bar{J}}^{[j]}
\Big{\{}-k_{R}\bm{e}^{[j]}_{\bm{R}}-k_{\Omega}\bm{e}^{[j]}_{\bm{\Omega}
}-\left([\bm{\Omega}]_{\times}\bm{R}^{\top}\bm{R}_{{\bm{c}}}\bm{\Omega}_{{\bm{c}}}\right)^{[j]}\\
&\!\!\!+\big{(}\bm{R}^{\top}\bm{R}_{{\bm{c}}}\bm{\dot{\Omega}}_{{\bm{c}}}\big{)}^{[j]}
-\bm{\bar{\phi}}_{\bm{\textit{R}}}^{[j]}+\underbrace{(\bm{J}^{-1}[\bm{\Omega}]_{\times}\bm{J}\bm{\Omega})^{[j]}}_{\text{if $\bm{J}$ is known}}\Big{\}},\\[-20pt]
\label{Md}
\end{aligned}
\end{equation}
where  vector $\bm{\mathcal{E}}_{\bm{\textit{x}} j}:= \left(\bm{e}_{\bm{x}}^{[j]}, \bm{e}_{\bm{v}}^{[j]}\right)^{\top}\!\!\!\!\in\mathbb{R}^2$ is the translational error vector along $\bm{\vec{e}}_j$-axis. The $\bm{\mathcal{K}}_{\bm{\textit{x}} j}:=\left(\bm{k_{\text{p}}}^{[j]},\bm{k_{\text{d}}}^{[j]}\right)^{\top}$ represents the gain vector for the translational Proportional-Derivative (PD) control along $\bm{\vec{e}}_j$-axis  with positive constants $\bm{k_{\text{p}}}$, $\bm{k_{\text{d}}}\in\mathbb{R}^{3}$. Symbol  $\bm{\delta}_{j3}$ denotes a Kronecker delta. The $k_{R}$ and $k_{\Omega}\in \mathbb{R}$ are positive gains for rotational PD control. In addition, in \textbf{\textit{Scenario 1 ($\bm{J}$ is known)}}, the term $(\bm{J}^{-1}[\bm{\Omega}]_{\times}\bm{J}\bm{\Omega})^{[j]}$ can be augmented as a compensation term. In \textbf{\textit{Scenario 2 ($\bm{J}$ is unknown)}}, $(\bm{J}^{-1}[\bm{\Omega}]_{\times}\bm{J}\bm{\Omega})^{[j]}$ is omitted and the neural networks intervene to learn and compensate for the total disturbance described in Eq.~\eqref{Dynamics with Augmented Disturbance2}, i.e., $\bm{\bar{\phi}}_{\bm{\textit{R}}}^{[j]}\to\bm{\phi}_{\bm{\textit{R}}}(\bm{J}, \bm{\Omega})^{[j]}$.

\textbf{\textit{Remark 1:}} In \textbf{\textit{Scenario 2}}, the neural networks must learn both internal and external disturbances, which may require more neurons and computational resources to reduce approximation errors.
\subsubsection{\textbf{SANM Design-Adaptive Laws}} Substituting the desired control wrench into system error dynamics, the adaptive laws for updating the estimated pseudo model $\bm{\bar{\Theta}}^{\text{rec}}$ are derived based on Lyapunov analysis in Appendix to ensure system stability.  The components of  the estimated pseudo model feature $\{\bm{\bar{m}}^{[j]},\bm{\bar{J}}^{[j]}\}_{1\leq j \leq 3}$ are updated online by following adaptive laws:
\begin{equation}
{
\footnotesize
\bm{\dot{\bar{m}}}^{[j]}\!\!:=\!\!
\begin{cases}
   \!\frac{-\bm{\bar{m}}^{[j]^2}}{\eta_{\textit{m}}}\bm{\mathcal{E}}_{\bm{\textit{x}} j}^{\top}\bm{P}_j\bm{B}\bm{\mathrm{F}_d}^{[j]}, \,\,\,\,\,\,\,\,\,\,\, \bm{\mathcal{E}}_{\bm{\textit{x}} j}^{\top}\bm{P}_j\bm{B}\bm{\mathrm{F}_d}^{[j]}>0\\[2pt]
 \!\frac{-\bm{\bar{m}}^{[j]^2}}{\eta_{\textit{m}}}\bm{\mathcal{E}}_{\bm{\textit{x}} j}^{\top}\bm{P}_j\bm{B}\bm{\mathrm{F}_d}^{[j]}, \,\,\,\,\,\,\,\,\,\,\,\bm{\mathcal{E}}_{\bm{\textit{x}} j}^{\top}\bm{P}_j\bm{B}\bm{\mathrm{F}_d}^{[j]}\leq 0, \,\,\bm{\bar{m}}^{[j]}\!<\!\overset{\tiny \text{max}}{m}\\[2pt]
 \mathfrak{s}_{\textit{m}}\frac{-\bm{\bar{m}}^{[j]^2}}{\eta_{\textit{m}}}, \,\, \,\,\,\,\,\,\,\,\,\,\,\,\,\,\,\,\,\,\,\,\,\,\,\,\,\,\,\,\,\,\,\,\,\bm{\mathcal{E}}_{\bm{\textit{x}} j}^{\top}\bm{P}_j\bm{B}\bm{\mathrm{F}_d}^{[j]}\leq 0 , \,\,\bm{\bar{m}}^{[j]}\!\geq\!\overset{\tiny \text{max}}{m}
\end{cases}
}
\label{Adaptive Law of mass}
\end{equation}
\begin{equation}
{
\footnotesize
\bm{\dot{\bar{\mathit{J}}}}^{[j]}\!\!:=\!\!
\begin{cases}
   \frac{-\bm{\bar{J}}^{[j]^2}}{\eta_{J}}\!\left(\!\bm{e}^{[j]}_{\bm{\Omega}}\!\!+\!c_R\bm{e}^{[j]}_{\bm{R}}\!\right)\!\bm{\mathrm{M}_d}^{[j]}\!, \,\, \left(\!\bm{e}^{[j]}_{\bm{\Omega}}\!\!+\!c_R\bm{e}^{[j]}_{\bm{R}}\!\right)\bm{\mathrm{M}_d}^{[j]}\!>\!0,\\[2pt]
 \frac{-\bm{\bar{J}}^{[j]^2}}{\eta_{J}}\!\left(\!\bm{e}^{[j]}_{\bm{\Omega}}\!\!+\!c_R\bm{e}^{[j]}_{\bm{R}}\!\right)\bm{\mathrm{M}_d}^{[j]}\!, \,\,\left(\!\bm{e}^{[j]}_{\bm{\Omega}}\!\!+\!c_R\bm{e}^{[j]}_{\bm{R}}\!\right)\!\bm{\mathrm{M}_d}^{[j]}\!\!\leq\! 0, \,\bm{\bar{J}}^{[j]}\!\!<\!\!\overset{\tiny \text{max}}{\bm{J}}\overset{[j]}{\rule{0pt}{1.5ex}}\\[2pt]
 \,\mathfrak{s}_{J}\frac{-\bm{\bar{J}}^{[j]^2}}{\eta_{J}}, \,\,\,\,\,\,\,\,\,\,\,\,\,\,\,\,\,\,\,\,\,\,\,\,\,\,\,\,\,\,\,\,\,\,\,\,\,\,\,\,\,\,\left(\!\bm{e}^{[j]}_{\bm{\Omega}}\!\!+\!c_R\bm{e}^{[j]}_{\bm{R}}\!\right)\!\bm{\mathrm{M}_d}^{[j]}\!\!\leq\! 0, \,\bm{\bar{J}}^{[j]}\!\!\geq\!\!\overset{\tiny \text{max}}{\bm{J}}\overset{[j]}{\rule{0pt}{1.5ex}}
\end{cases}
}
\label{Adaptive Law of Inertia Tensor}
\end{equation}
where $\eta_{\textit{m}}\!\in\!\mathbb{R}$, $\eta_{J}\!\in\!\mathbb{R}$ and $c_R\!\in\!\mathbb{R}$ are positive constants. The constant $\overset{\tiny \text{max}}{m}\in\mathbb{R}$ is the preset maximum mass and  $\overset{\tiny \text{max}}{\bm{J}}\overset{[j]}{\rule{0pt}{1.5ex}}\in\mathbb{R}$ denotes the maximum moment of inertia along $\bm{\vec{b}}_{j}$-axis. $\mathfrak{s}_{\textit{m}}\in\mathbb{R}$ and $\mathfrak{s}_{J}\in\mathbb{R}$ are scaling factors. The $\bm{P}_j\in\mathbb{R}^{2\times2}$ denotes $j^{th}$ Lyapunov matrix  and  $\bm{B}=\left(0,1\right)^{\top}$ is a unit basis vector.

The estimation errors of pseudo model
$\{\widetilde{m}_j,\widetilde{J}_{j}\}_{1\leq j \leq3}$ are defined in a reciprocal form: 
\begin{align}
     \widetilde{m}_j& \triangleq\frac{1}{m}-\frac{1}{\bm{\bar{m}}^{[j]}},\,\,\,
      \widetilde{J}_{j}\triangleq\frac{1}{\bm{J}^{[j]}}-\frac{1}{\bar{\bm{J}}^{[j]}},
\label{Estimation errors of pseudo model}
\end{align}
where $\widetilde{m}_j$ denotes the estimation error of mass along  $\bm{\vec{e}}_{j}$-axis and  $\widetilde{J}_{j}$ denotes the estimation error of  the moment of inertia along  $\bm{\vec{b}}_{j}$-axis.

\subsubsection{\textbf{SANM Design-Shallow Neural Networks (SNN)}}
\begin{figure}[!t]
      \centering
      \includegraphics[scale=0.18]{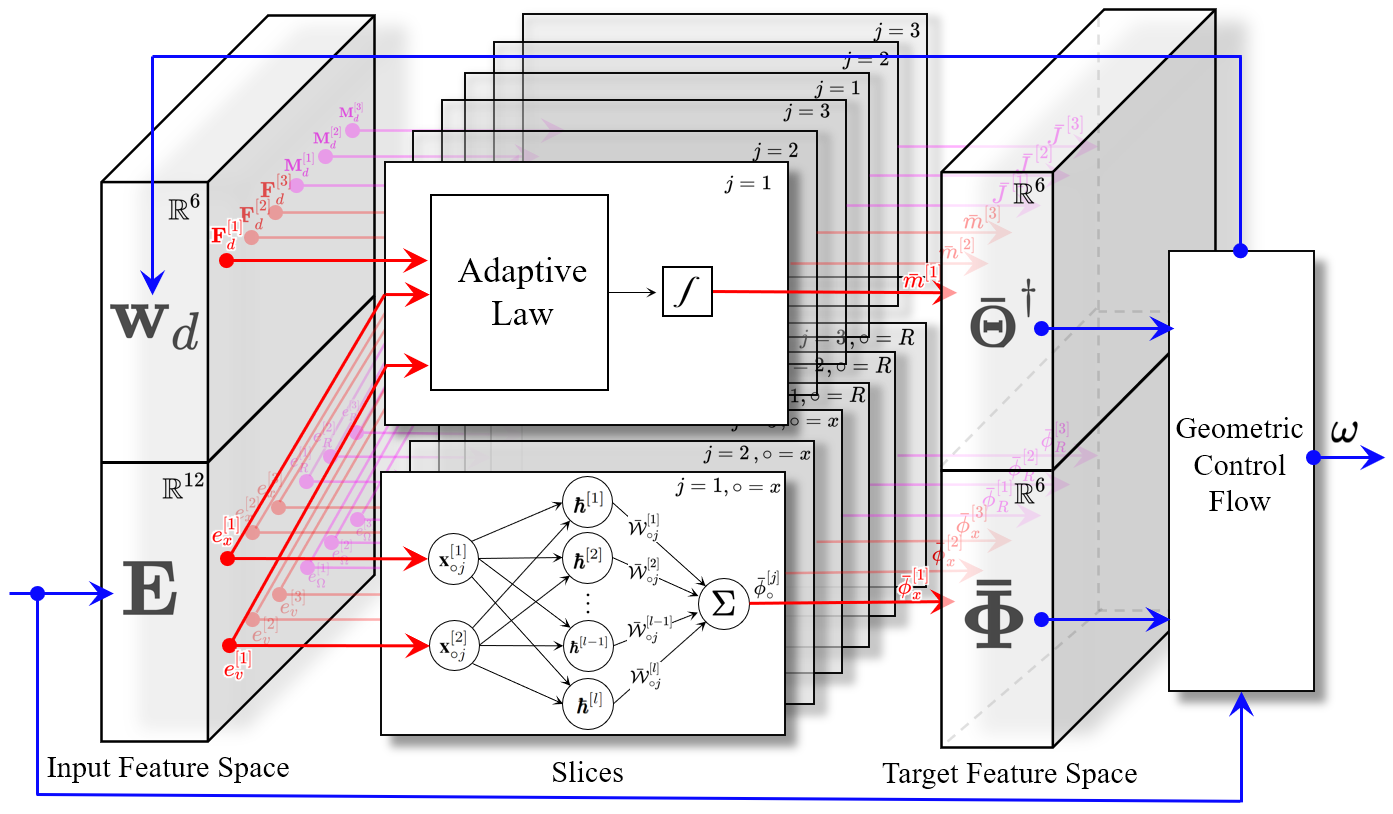}
      \caption{\footnotesize The  structure of Sliced Adaptive-Neuro Mapping (SANM). The 12 slices Correspond to 12-dimensional state, respectivly. }
      \label{Gao2}
\end{figure}
Based on the universal approximation theorem \cite{1989 Multilayer feedforward networks are universal approximators}, the aforementioned disturbance dynamics mapping $\bm{\Phi}=\left(\bm{\phi}^{\top}_{\bm{\textit{x}}}, \bm{\phi}^{\top}_{\bm{\textit{R}}}\right)^{\top}=\bm{\mathcal{S}}^\dagger_{\Phi}(\mathbf{E}):\bm{\mathcal{C}}\to\mathbb{R}^6$ can be approximated  on a compact domain $\bm{\mathcal{C}}\subset\mathbb{R}^{12}$ by multiple neural networks with sufficient capacity. For each $j^{th}$ component of $\bm{\phi}_{\bm{\textit{x}}}$ and $\bm{\phi}_{\bm{\textit{R}}}$, a Radial Basis Function (RBF) neural network with 2 inputs-$l$ hidden layer neurons-1 output (2-$l$-1) structure (see Fig.~\ref{Gao2}) is deployed as follows:
\begin{equation}
\bm{\phi}_{\circ}^{[j]}=\bm{\mathcal{W}}_{\circ j}^{\top}\bm{\hbar}(\textbf{x}_{\circ j})+\epsilon_{\circ j},
\label{Phi}
\end{equation}
where subscript $\bullet_{\circ\in\{\bm{\textit{x}}, \bm{\textit{R}}\}}$ indicates terms associated with translational and rotational dynamics, respectively. 
The $\textbf{x}_{\circ j}\in \mathbb{R}^2$ denotes the input vector of $j^{th}$ neural network, and $\bm{\mathcal{W}}_{\circ j}\in\mathbb{W}_{\circ j}$ represent the corresponding weights vector bounded within a compact set $\mathbb{W}_{\circ j}=\{\bm{\mathcal{W}}_{\circ j}\in\mathbb{R}^l\,|\,\|\bm{\mathcal{W}}_{\circ j}\|\leq r_w\}$ for a positive constant $r_w$. The $\bm{\hbar}(\textbf{x}_{\circ j})\in\mathbb{R}^l$ denotes the Gaussian activation function and the $\epsilon_{\circ j}\in\mathbb{R}^+$ represents an arbitrarily small intrinsic approximation error, i.e.,  $\epsilon_{\circ j}\to 0^+$. 

The output of $k^{th}$ hidden layer neurons is expressed as:
\begin{equation}
\begin{aligned}
&\bm{\hbar}^{[k]}(\textbf{x}_{\circ j}):=\mathrm{exp}\left(-\frac{\lVert\textbf{x}_{\circ j}-\textbf{c}_{k}\rVert^2}{2b^{2}_k}\right),
\end{aligned}
\label{Gaussian activation function}
\end{equation}
where $\textbf{c}_{k}\in\mathbb{R}^2$ denotes the center vector of $k^{th}$ neurons and $b_k\in\mathbb{R}$ denotes the width of $k^{th}$ Gaussian function, $1\leq k \leq l$. 

To approximate Eq.~\eqref{Phi}, the estimated disturbance dynamics $\bm{\bar{\phi}}_{\circ}^{[j]}$ is represented by a neural network with time-varying estimated weights $\bm{\bar{\mathcal{W}}}_{\circ j}\in\mathbb{W}_{\circ j}$. This approximation is expressed as follows:
\begin{equation}
\begin{aligned}
\bm{\bar{\phi}}_{\circ}^{[j]}:=\bm{\bar{\mathcal{W}}}_{\circ j}^{\top}\bm{\hbar}(\textbf{x}_{\circ j}),
\end{aligned}
\label{Phi_hat}
\end{equation}
where input  $\textbf{x}_{\bm{\textit{x}} j}:=\bm{\mathcal{E}}_{\bm{\textit{x}} j}$ takes  the translational error vector along $\bm{\vec{e}}_j$-axis, and input $\textbf{x}_{\bm{\textit{R}} j}:= \left(\bm{e}_{\bm{R}}^{[j]}, \bm{e}_{\bm{\Omega}}^{[j]}\right)^{\top}$ takes the rotational error vector along $\bm{\vec{b}}_j$-axis. 

According to the Lyapunov analysis in Appendix,  the estimated weights $\bm{\bar{\mathcal{W}}}_{\bm{\textit{x}} j}\in\mathbb{W}_{\bm{\textit{x}} j}$ and $\bm{\bar{\mathcal{W}}}_{\bm{\textit{R}} j}\in\mathbb{W}_{\bm{\textit{R}} j}$ are designed to be updated online by the following Lyapunov adaptation:
\begin{align}
\bm{\dot{\bar{\mathcal{W}}}}_{\bm{\textit{x}} 
j}:=&\gamma_{\bm{\textit{x}}j}\bm{\mathcal{E}}_{\bm{\textit{x}} j}^{\top}\bm{P}_j\bm{B}\bm{\hbar}(\textbf{x}_{\bm{\textit{x}} j}),
\label{Estimated Weights_x}\\
\bm{\dot{\bar{\mathcal{W}}}}_{\bm{\textit{R}} 
 j}:=&\gamma_{\bm{\textit{R}}j} \left(\bm{e}^{[j]}_{\bm{\Omega}}+c_R\bm{e}^{[j]}_{\bm{R}}\right)\bm{\hbar}(\textbf{x}_{\bm{\textit{R}} j}),
\label{Estimated Weights_R}
\end{align}
where $\gamma_{\bm{\textit{x}}j}$, $\gamma_{\bm{\textit{R}}j}$ and $c_R$ are corresponding  positive constants. The optimal weights can be identified by these laws are expressed as:
\begin{equation}
    \bm{\mathcal{W}}_{\circ j}^*\triangleq\mathrm{arg}\, \underset{\bm{\mathcal{W}}_{\circ j}\in\mathbb{W}_{\circ j}}{\mathrm{min}}\left(\mathrm{sup}\big{\vert}\bm{\phi}_{\circ }^{[j]}-\bm{\bar{\phi}}_{\circ }^{[j]}\big{\vert}\right), \label{W*}
\end{equation}
where $\mathrm{arg}\,\mathrm{min}$ denotes the value of $ \bm{\mathcal{W}}_{\circ j}$ that minimizes the supremum of the error between $\bm{\phi}_{\circ}^{[j]}$ and $\bm{\bar{\phi}}_{\circ}^{[j]}$.

The optimal approximation error is then defined as follows:
\begin{equation}
     \bm{\varpi}^{[j]}_{\circ}\triangleq\bm{\phi}_{\circ}^{[j]}-\bm{\bar{\phi}}_{\circ}^{[j]}(\textbf{x}_{\circ j}\vert\bm{\mathcal{W}}^*_{\circ j}),
     \label{varpi}
\end{equation}
where $\|\bm{\varpi}^{[j]}_{\circ}\|$ is bounded according to the universal approximation theorem \cite{1989 Multilayer feedforward networks are universal approximators} and Proposition 0. Here, $\bm{\varpi}^{[j]}_{\circ}\in\mathbb{R}$ denotes the $j^{th}$ component of the optimal approximation error vector $\bm{\varpi}_{\circ}\in\mathbb{R}^{3}$.

From Eqs.~\eqref{Phi_hat}, \eqref{W*}, \eqref{varpi}, the problem of the approximation error $\bm{\phi}_{\circ }^{[j]}-\bm{\bar{\phi}}_{\circ }^{[j]}$ can be transformed into the problem of weight estimation error:
\begin{equation}
   \bm{\phi}_{\circ}^{[j]}-\bm{\bar{\phi}}_{\circ }^{[j]}=\bm{\tilde{\mathcal{W}}}_{\circ j}^{\top}\bm{\hbar}(\textbf{x}_{\circ j})+\bm{\varpi}^{[j]}_{\circ},
    \label{approximation error to weight error}
\end{equation}
where the weight estimation error is defined as:
\begin{equation}
    \bm{\tilde{\mathcal{W}}}_{\circ j}\triangleq\bm{\mathcal{W}}^*_{\circ j}-\bm{\bar{\mathcal{W}}}_{\circ j}.
    \label{weight error}
\end{equation}
\subsection{Propositions}
To satisfy the prerequisite of the universal approximation theorem, we first establish the following proposition:

\textbf{\textit{Proposition 0: (Compact Set Constraint on Neural Network Inputs)}} Full-State error $\mathbf{E}$ is bounded by a compact set: $ \bm{\mathcal{C}}=\Big{\{}\exists r_c > 0,\mathbf{E}\in\mathbb{R}^{12}| \,\|\mathbf{E}\|\leq\!\|\bm{e}_{\bm{x}}\|\!+\!\|\bm{e}_{\bm{v}}\| \!+\!\|\bm{e}_{\bm{R}}\|\! +\!\|\bm{e}_{\bm{\Omega}}\|\leq r_{c}\Big{\}}$
for a positive constant $r_{c}$. This implies that all the neural networks inputs $\{\textbf{x}_{\circ j}\}_{1 \leq j \leq 3,\,\circ\in\{\bm{\textit{x}}, \bm{\textit{R}}\}}$ are also bounded within their respective compact sets. 

Then, we consider the following almost global domain of attraction for the initial conditions of rotational dynamics:
\begin{equation}
{\small
   \begin{aligned}
    \mathcal{D}_{{\bm{\textit{R}}0}}\!=&\Big{\{}
\,\,\,0<\Psi_{\textit{R}}\big{(}\bm{R}(0),\bm{R}_{\bm{c}}(0)\big{)}<2,\\[-0pt]
&\,\,\,\,\,\,\|\bm{e_R}(0)\|=\sqrt{\Psi_{\textit{R}}(0)\big{(}2-\Psi_{\textit{R}}(0)\big{)}},\\
&\,\,\,\,\,\,\|\bm{e_\Omega}(0)\|^2<2k_R\Big{(}2-\Psi_{\textit{R}}\big{(}\bm{R}(0),\bm{R}_{\bm{c}}(0)\big{)}\Big{)}\,\,\,\, \Big{\}},
\end{aligned} 
}
\end{equation}
where $\Psi_{\textit{R}}\!:\!\mathbf{SO}(3)\times\mathbf{SO}(3)\!\to\!\mathbb{R}$ denotes an attitude configuration error scalar function as noted in Eqs.~\eqref{Psi_R} and \eqref{Psi_R_bound}. When $0\!<\!\Psi_{\textit{R}}\!<\!2$, it covers almost $\mathbf{SO}(3)$, except for singular points corresponding to a rotation of exactly $180^\circ$.
This domain of attraction differs from that in \cite{2010 Geometric tracking control of a quadrotor UAV on SE(3)} as its size is independent of the inertia tensor $\bm{J}$ due to the adaptive nature of SANM. Within this domain, we show the Near-Exponential Stability (NES) of the rotational dynamics.

\textbf{\textit{Definition 1: (Near-Exponentially Stable)}} The solution $z(t)$ of a dynamic system is \textit{Near-Exponentially Stable} around $z\!=\!0$ if there exist positive constants $\alpha$, $\beta$, and $\mathfrak{d}$ such that if $\|z(0)\|\leq\mathfrak{d}$, then $\|z(t)\|\leq\alpha\|z(0)\|e^{-\beta t}+\epsilon$, $t\geq0$. Here, $\epsilon\to0^+$ denotes an arbitrarily small positive bound.

Near-Exponential Stability (NES) is a stronger notion of  Practical Exponential Stability (PES). While PES ensures exponential convergence to a ball around the zero equilibrium with a bounded radius, NES allows this radius to be made arbitrarily small to approach exponential stability. 

\textbf{\textit{Proposition 1: (Almost Global Near-Exponential Stability of Rotational Error Dynamics under Dynamic Disturbance Moments and Inertia Uncertainties)}} Under the attitude control compensated by SANM on $\mathbf{SO}(3)$ and the initial condition that $\bm{z}_{\bm{\textit{R}}}(0)\!\!\in\!\mathcal{D}_{\bm{\textit{R}}0}$, the solution of the rotational error dynamics associated with $\bm{e}_{\bm{R}},\bm{e}_{\bm{\Omega}}$ is almost globally near-exponentially stable around zero equilibrium, and the estimation errors $\{ \widetilde{J}_{j}, \bm{\tilde{\mathcal{W}}}_{\bm{\textit{R}} j}\}_{1\leq j \leq3}$ remain uniformly bounded. This result holds despite the presence of time-varying disturbance moments and an unknown inertia tensor.

Next, we consider the following local domain of attraction for complete dynamics:
\begin{equation}
{\small
   \begin{aligned}
    \mathcal{D}_{0}\!=\Big{\{}
&\,\,\,\, \|\bm{\mathcal{E}}_{\bm{\textit{x}} j}(0)\|<\overset{\tiny \text{max}}{\mathcal{E}_{\bm{\textit{x}} j}}, \,0<\Psi_{\textit{R}}\big{(}\bm{R}(0),\bm{R}_{\bm{c}}(0)\big{)}<1,\\[-0pt]
&\,\,\,\,\|\bm{e_R}(0)\|=\sqrt{\Psi_{\textit{R}}(0)\big{(}2-\Psi_{\textit{R}}(0)\big{)}},\\
&\,\,\,\,\|\bm{e_\Omega}(0)\|^2<2k_R\Big{(}1-\Psi_{\textit{R}}\big{(}\bm{R}(0),\bm{R}_{\bm{c}}(0)\big{)}\Big{)}\,\,\,\, \Big{\}},
\end{aligned} 
}
\end{equation}
where $\overset{\tiny \text{max}}{\mathcal{E}_{\bm{\textit{x}} j}}$ denotes the upper bound of $\|\bm{\mathcal{E}}_{\bm{\textit{x}} j}\|$. When $0\!<\!\Psi_{\textit{R}}\!<\!1$, this implies that the attitude error angle does not exceed $90^\circ$.

\textbf{\textit{Proposition 2: (Local Near-Exponential Stability of Complete Dynamics Under Complex Dynamic Disturbances and Model Uncertainties)}}  Under the position-attitude coupled control compensated by SANM on $\mathbf{SE}(3)$ and initial condition that $\bm{z}(0)\!\in\!\mathcal{D}_0$, the solution of all state errors $\bm{e}_{\bm{x}},\bm{e}_{\bm{v}}, \bm{e}_{\bm{R}},\bm{e}_{\bm{\Omega}}$ is locally near-exponentially stable around zero equilibrium within $\mathcal{D}_{0}$, and all estimation errors $\{\widetilde{m}_j, \widetilde{J}_{j},  \bm{\tilde{\mathcal{W}}}_{\bm{\textit{x}} j}, \bm{\tilde{\mathcal{W}}}_{\bm{\textit{R}} j}\}_{1\leq j \leq3}$ remain uniformly bounded. This result holds despite the presence of time-varying multi-dimentional disturbances (forces and moments) and model uncertainties (unknown inertia tensor and mass).

\textbf{\textit{Proofs of Propositions:}} See Appendix.

\section{Conclusion}
\label{Conclusion}
In this paper, we introduced Dimension-Decomposed Learning (DiD-L) and demonstrated the feasibility of its instance, Sliced Adaptive-Neuro Mapping (SANM). The exponential convergence of all quadrotor state errors to an arbitrarily small ball has been ensured, but it also poses new open questions: \textit{(1) How can we further reduce the size of the ball? Is increasing the number of neurons in the SNN sufficient?} \textit{(2) Can the idea of dimension decomposition be applied to offline training in data-driven control?} Future work will focus on addressing these open questions.

\appendix
The subsequent Lyapunov analysis is conducted in the following open domain:
\begin{equation}
   \begin{aligned}
    \mathcal{D}\!=&\Big{\{}\Big{(}\bm{e_x},\bm{e_v},\bm{e_R}, \bm{e_\Omega},(\tilde{m}_j,\widetilde{J}_{j},\bm{\tilde{\mathcal{W}}}_{\bm{\textit{x}} j},\bm{\tilde{\mathcal{W}}}_{\bm{\textit{R}} j})_{1\leq j \leq3}\Big{)}\!\in\mathbb{R}^3\!\times\!\mathbb{R}^3\\[-5pt]
&\!\times\!\mathbb{R}^3\!\times\!\mathbb{R}^3\!\times\!\prod^{3}_{j=1}(\mathbb{R}\!\times\!\mathbb{R}\!\times\!\mathbb{R}^{l}\!\times\!\mathbb{R}^{l})\big{|} \,\|\bm{e_x}\| \!+\!\|\bm{e_v}\|\!+\!\|\bm{e_R}\|\\[-10pt]
&\!+\!\|\bm{e_\Omega}\|\!+\!\sum_{j=1}^3(\|\tilde{m}_j\|\!+\!\|\widetilde{J}_{j}\|\!+\!\|\bm{\tilde{\mathcal{W}}}_{\bm{\textit{x}} j}\|\!+\!\|\bm{\tilde{\mathcal{W}}}_{\bm{\textit{R}} j}\|) < r_d\Big{\}}
\end{aligned} 
\label{D}
\end{equation}
for a positive constant $r_d$. In this domain, the bounded desired trajectory 
$\bm{x}_{{\bm{d}}}(t)$ is carefully selected to prevent any violation of the compact set constraint established in Proposition 0. The $\|\bm{e_R}\|$ is bounded by $\|\bm{e_R}\|=\sqrt{\Psi_{\textit{R}}(2-\Psi_{\textit{R}})}\leq\sqrt{\psi_{\textit{R}}(2-\psi_{\textit{R}})}<1$ with a positive scalar $0<\psi_{\textit{R}}<2$ and an attitude configuration error scalar function proposed in \cite{2010 Geometric tracking control of a quadrotor UAV on SE(3)}:

\begin{equation}
    \Psi_{\textit{R}}(\bm{R},\bm{R}_{\bm{c}})\triangleq\frac{1}{2}\mathrm{tr}\Big{[}\mathrm{I}^{3\times3}-\bm{R}_{\bm{c}}^{\top}\bm{R}\Big{]},
    \label{Psi_R}
\end{equation}
where $\Psi_{\textit{R}}:\mathbf{SO}(3)\times\mathbf{SO}(3)\to\mathbb{R}$ is positive definite and constrained by:
\begin{equation}
    \frac{1}{2}\|\bm{e_R}\|^{2}\leq\Psi_{\textit{R}}\leq\frac{1}{2-\psi_{\textit{R}}}\|\bm{e_R}\|^{2}.
    \label{Psi_R_bound}
\end{equation}

\subsection{Error Dynamics}
\label{Appendix_Error Dynamics}
\subsubsection{Translational Error Dynamics}
 By taking the time derivative of the translational error terms from Eq.~\eqref{errors}, the error dynamics is given as follows:
\begin{align}
    \bm{\dot{e}_x}&=\bm{e_v},\label{dot_e_x}\\
    \bm{\dot{e}_v}&=-\frac{1}{m}f\bm{R}\bm{\vec{e}}_3+g\bm{\vec{e}}_3+\bm{\phi}_{\bm{\textit{x}}}-\bm{\ddot{x}_d}.\label{dot_e_v}
\end{align}
Following the formulation in \cite{2010 Geometric tracking control of a quadrotor UAV on SE(3)}, we define $\mathcal{X}\in\mathbb{R}^{3}$ as:
\begin{equation}
    \mathcal{X}\equiv\frac{f_d}{\bm{\vec{e}}_3^{\top}\bm{R}_{\bm{c}}^{\top}\bm{R}\bm{\vec{e}}_3}\Big{\{}\left(\bm{\vec{e}}_3^{\top}\bm{R}_{\bm{c}}^{\top}\bm{R}\bm{\vec{e}}_3\right)\bm{R}\bm{\vec{e}}_3-\bm{R}_{\bm{c}}\bm{\vec{e}}_3\Big{\}}
    \label{X}
\end{equation}
where $0<\bm{\vec{e}}_3^{\top}\bm{R}_{\bm{c}}^{\top}\bm{R}\bm{\vec{e}}_3<1$ and $\mathcal{X}^{[j]}\geq0$ for $1\leq j \leq 3$. Substituting Eqs.~\eqref{mapping deviations} and \eqref{X} into Eq.~\eqref{dot_e_v}, the following expression is obtained:
\begin{equation}
   \bm{\dot{e}_v}\!=\!\frac{1}{m}\left(-\frac{f_d}{\bm{\vec{e}}_3^{\top}\bm{R}_{\bm{c}}^{\top}\bm{R}\bm{\vec{e}}_3}\bm{R}_{\bm{c}}\bm{\vec{e}}_3\!-\!\Delta_f \bm{R}\bm{\vec{e}}_3\!-\!\mathcal{X}\right)\!+\!g\bm{\vec{e}}_3\!+\!\bm{\phi}_{\bm{\textit{x}}}\!-\!\bm{\ddot{x}_d}.
\end{equation}
Given that $\bm{R}_{\bm{c}}\bm{\vec{e}}_3=\bm{\vec{b}}_{3\bm{c}}:= -\bm{\mathrm{F}_d}/\|\bm{\mathrm{F}_d}\|$ and $f_d:=-\bm{\mathrm{F}_d} \cdot \bm{R}\bm{\vec{e}}_3$, we can derive $f_d:=(\|\bm{\mathrm{F}_d}\|\bm{R}_{\bm{c}}\bm{\vec{e}}_3)\cdot\bm{R}\bm{\vec{e}}_3$ and it follows that:
\begin{equation}
    -\frac{f_d}{\bm{\vec{e}}_3^{\top}\bm{R}_{\bm{c}}^{\top}\bm{R}\bm{\vec{e}}_3}\bm{R}_{\bm{c}}\bm{\vec{e}}_3=\bm{\mathrm{F}_d}.
\end{equation}
Therefore, the velocity error dynamics is rewritten as:
\begin{equation}
    \bm{\dot{e}_v}=\frac{1}{m}\bm{\mathrm{F}_d}+g\bm{\vec{e}}_3+\bm{\phi}_{\bm{\textit{x}}}-\bm{\ddot{x}_d}-\frac{1}{m}\left(\Delta_f \bm{R}\bm{\vec{e}}_3+\mathcal{X}\right).
\end{equation}
Substituting Eqs.~\eqref{Fd} and \eqref{Estimation errors of pseudo model}, the velocity error dynamics along $\bm{\vec{e}}_j$-axis is rearranged as:
\begin{equation}
\bm{\dot{e}_v}^{[j]}\!=\!-\bm{\mathcal{K}}_{\bm{\textit{x}} j}^{\top}\bm{\mathcal{E}}_{\bm{\textit{x}} j}+\widetilde{m}_j\bm{\mathrm{F}_d}^{[j]}+\left(\bm{\phi}_{\bm{\textit{x}}}^{[j]}-\bm{\bar{\phi}}_{\bm{\textit{x}}}^{[j]}\right)-\frac{1}{m}\left(\Delta_f \bm{R}\bm{\vec{e}}_3\!+\!\mathcal{X}\right)^{[j]}.
\end{equation}

Introducing Eq.~\eqref{dot_e_x}, the translational error dynamics along $\bm{\vec{e}}_j$-axis can be further expressed as:
\begin{equation} 
 {\small
\begin{aligned}
      \bm{\dot{\mathcal{E}}}_{\bm{\textit{x}} j}={\renewcommand{\arraystretch}{1}\begin{bmatrix}\bm{\dot{e}}_{\bm{x}}^{[j]}\\ \bm{\dot{e}}_{\bm{v}}^{[j]}\\\end{bmatrix}}=&\bm{\Lambda}_{\bm{\textit{x}} j}\bm{\mathcal{E}}_{\bm{\textit{x}} j}+\bm{B}\bigg{\{} \widetilde{m}_{j}\bm{\mathrm{F}_d}^{[j]}+\left(\bm{\phi}_{\bm{\textit{x}}}^{[j]}-\bm{\bar{\phi}}_{\bm{\textit{x}}}^{[j]}\right)
      \\&-\frac{1}{m}\left(\Delta_f \bm{R}\bm{\vec{e}}_3\right)^{[j]}-\frac{1}{m}\mathcal{X}^{[j]}\bigg{\}},
\end{aligned}
}
\label{translational error dynamics1}
\end{equation}
where
\begin{equation}
    \bm{\Lambda}_{\bm{\textit{x}} j}={\renewcommand{\arraystretch}{1}\begin{bmatrix}0&1\\-\bm{k_{\text{p}}}^{[j]}&-\bm{k_{\text{d}}}^{[j]}\end{bmatrix}}, \,\,\,\,\, \bm{B}={\renewcommand{\arraystretch}{1}\begin{bmatrix}\,0\,\,\\\,1\,\,\end{bmatrix}}.
\end{equation}
From Eq.~\eqref{approximation error to weight error}, the problem of the approximation error $\bm{\phi}_{\bm{\textit{x}}}^{[j]}-\bm{\bar{\phi}}_{\bm{\textit{x}}}^{[j]}$ can be transformed into the problem of weight estimation error. Therefore, the translational error dynamics can ultimately be given by:
 \begin{align}
&\begin{aligned}
      \bm{\dot{\mathcal{E}}}_{\bm{\textit{x}} j}={\renewcommand{\arraystretch}{1}\begin{bmatrix}\bm{\dot{e}}_{\bm{x}}^{[j]}\\ \bm{\dot{e}}_{\bm{v}}^{[j]}\\\end{bmatrix}}=\bm{\Lambda}_{\bm{\textit{x}} j}\bm{\mathcal{E}}_{\bm{\textit{x}} j}+\bm{B}\bigg{\{} \widetilde{m}_j\bm{\mathrm{F}_d}^{[j]}+ \bm{\tilde{\mathcal{W}}}_{\bm{\textit{x}} j}^{\top}\bm{\hbar}(\textbf{x}_{\bm{\textit{x}} j})+\bm{\varpi}^{[j]}_{\bm{\textit{x}}}
\end{aligned}\notag\\
&\begin{aligned}
\,\,\,\,\,\,\,\,\,\,\,\,\,\,\,\,\,\,\,\,\,\,\,\,\,\,\,\,\,\,\,\,\,\,\,\,\,\,\,\,-\frac{1}{m}\left(\Delta_f \bm{R}\bm{\vec{e}}_3\right)^{[j]}-\frac{1}{m}\mathcal{X}^{[j]}\bigg{\}}.
\end{aligned}
\label{translational error dynamics2}
\end{align}
For subsequent derivations, we denote the second term in Eq.~\eqref{translational error dynamics2} as $G_j$ for simplicity:
 \begin{align}
\begin{aligned}
      \bm{\dot{\mathcal{E}}}_{\bm{\textit{x}} j}=\bm{\Lambda}_{\bm{\textit{x}} j}\bm{\mathcal{E}}_{\bm{\textit{x}} j}+G_j.
\end{aligned}
\label{translational error dynamics3}
\end{align}
\subsubsection{Rotational Error Dynamics}
\label{Rotational Error Dynamics}
Following the formulation in \cite{2010 Geometric tracking control of a quadrotor UAV on SE(3)} and \cite{2011 Geometric tracking control of the attitude dynamics of a rigid body on SO(3)}, the rotational error dynamics is given by:
\begin{align}
\bm{\dot{e}_R}&=\frac{1}{2}\left(\bm{R_c}^{\top}\bm{R}[\bm{e_{\Omega}}]_{\times}+[\bm{e_{\Omega}}]_{\times}\bm{R}^{\top}\bm{R_c}\right)^{\vee}\notag\\
&=\frac{1}{2}\left(\mathrm{tr}\big{[}\bm{R}^{\top}\bm{R_c}\big{]}\mathrm{I}^{3\times3}-\bm{R}^{\top}\bm{R_c}\right)\bm{e_{\Omega}}\notag\\
&\equiv Y(\bm{R_c}^{\top}\bm{R})\bm{e_{\Omega}},\label{rotational error dynamics1}\\
\bm{\dot{e}}_{\bm{\Omega}}&=\bm{\dot{\Omega}}+[\bm{\Omega}]_{\times}\bm{R}^{\top}\bm{R_c}\bm{\Omega_c}-\bm{R}^{\top}\bm{R_c}\bm{\dot{\Omega}_c},
\label{rotational error dynamics2}
\end{align}
where $\|Y(\bm{R_c}^{\top}\bm{R})\|\leq1$ for any $\bm{R_c}^{\top}\bm{R}\in\mathbf{SO}(3)$.

In \textbf{\textit{Scenario 1 ($\bm{J}$ is known)}}, substituting Eqs.~\eqref{Dynamics with Augmented Disturbance} and \eqref{mapping deviations} into Eq.~\eqref{rotational error dynamics2}, we have:
\begin{align}
\bm{\dot{e}_{\Omega}}=&\bm{J}^{-1}\left(\bm{\mathrm{M}_d}-[\bm{\Omega}]_{\times}\bm{J}\bm{\Omega}+\bm{\Delta_{\mathrm{M}}}\right)+\bm{\phi}_{\bm{\textit{R}}}\notag\\
    &+[\bm{\Omega}]_{\times}\bm{R}^{\top}\bm{R_c}\bm{\Omega_c}-\bm{R}^{\top}\bm{R_c}\bm{\dot{\Omega}_c}.
    \label{rotational error dynamics3}
\end{align}
Further substituting Eqs.~\eqref{Md} and \eqref{Estimation errors of pseudo model}, the augular velocity error dynamics along $\bm{\vec{b}}_{j}$-axis is rearranged as:
\begin{equation}
{\footnotesize
    \begin{aligned}
    \bm{\dot{e}_{\Omega}}^{[j]}=&-k_{R}\bm{e}^{[j]}_{\bm{R}}-k_{\Omega}\bm{e}^{[j]}_{\bm{\Omega}
}+\widetilde{J}_{j}\bm{\mathrm{M}_d}^{[j]}+\left(\bm{\phi}_{\bm{\textit{R}}}^{[j]}-\bm{\bar{\phi}}_{\bm{\textit{R}}}^{[j]}\right)\\
    &\!\!\!\!+\underbrace{(\bm{J}^{-1}[\bm{\Omega}]_{\times}\bm{J}\bm{\Omega})^{[j]}}_{\text{if $\bm{J}$ is known}}-\left(\bm{J}^{-1}[\bm{\Omega}]_{\times}\bm{J}\bm{\Omega}\right)^{[j]}+\left(\bm{J}^{-1}\bm{\Delta_{\mathrm{M}}}\right)^{[j]},\\[-5pt]
\end{aligned}
}
\label{rotational error dynamics4}
\end{equation}
with the fact that $\bm{J}$ is always diagonalizable such that $\bm{J}^{-1[j]}=1/{\bm{J}^{[j]}}$. From Eq.~\eqref{approximation error to weight error}, the problem of the approximation error $\bm{\phi}_{\bm{\textit{R}}}^{[j]}-\bm{\bar{\phi}}_{\bm{\textit{R}}}^{[j]}$ is transformed into the problem of weight estimation error. Therefore, the rotational error dynamics can ultimately be given by:
\begin{equation}
{\footnotesize
    \begin{aligned}
    \bm{\dot{e}_{\Omega}}^{[j]}=&-k_{R}\bm{e}^{[j]}_{\bm{R}}-k_{\Omega}\bm{e}^{[j]}_{\bm{\Omega}
}+\widetilde{J}_{j}\bm{\mathrm{M}_d}^{[j]}+\bm{\tilde{\mathcal{W}}}_{\bm{\textit{R}} j}^{\top}\bm{\hbar}(\textbf{x}_{\bm{\textit{R}} j})+\bm{\varpi}^{[j]}_{\bm{\textit{R}}}\\
    &\!\!\!\!+\underbrace{(\bm{J}^{-1}[\bm{\Omega}]_{\times}\bm{J}\bm{\Omega})^{[j]}}_{\text{if $\bm{J}$ is known}}-\left(\bm{J}^{-1}[\bm{\Omega}]_{\times}\bm{J}\bm{\Omega}\right)^{[j]}+\left(\bm{J}^{-1}\bm{\Delta_{\mathrm{M}}}\right)^{[j]},\\[-5pt]
\end{aligned}
}
\label{rotational error dynamics5}
\end{equation}
where, if the knowledge of inertia tensor $\bm{J}$ is augmented in Eq.~\eqref{Md}, the sixth term appears to cancel out the seventh term.

In \textbf{\textit{Scenario 2 ($\bm{J}$ is unknown)}}, since the neural networks intervene to learn and compensate for the internal disturbance term $\bm{J}^{-1}[\bm{\Omega}]_{\times}\bm{J}\bm{\Omega}$, by submitting
Eq.~\eqref{Dynamics with Augmented Disturbance2} into Eq.~\eqref{rotational error dynamics2} instead of Eq.~\eqref{Dynamics with Augmented Disturbance}, all the terms associated with $\bm{J}^{-1}[\bm{\Omega}]_{\times}\bm{J}\bm{\Omega}$ in Eqs.~\eqref{rotational error dynamics3} \eqref{rotational error dynamics4} \eqref{rotational error dynamics5} vanish. Note that the approximation errors in this scenario may increase if the representation ability of the neural networks is insufficient. 
\subsection{Lyapunov Candidate}
\subsubsection{Candidate for Attitude State Error }
Define the  Lyapunov candidate function for the attitude state error as:
\begin{equation}
{\small
\begin{aligned}
\bm{\mathcal{V}_{\bm{\textit{R},s}}}=k_{R}\Psi_{\textit{R}}+\bm{\sum}_{j=1}^3\Big{(}\frac{1}{2}\|\bm{e}^{[j]}_{\bm{\Omega}}\|^2+c_R\bm{e}^{[j]}_{\bm{R}}\bm{e}^{[j]}_{\bm{\Omega}}\Big{)}
\end{aligned}
}
\label{V_R,s}
\end{equation}
where $k_R$ and $c_R$ are positive constants. The compact form of Eq.~\eqref{V_R,s} can be expressed as:
\begin{equation}
{\small
\begin{aligned}
\bm{\mathcal{V}_{\bm{\textit{R},s}}}=k_{R}\Psi_{\textit{R}}+\frac{1}{2}\|\bm{e}_{\bm{\Omega}}\|^2+c_R\bm{e}_{\bm{R}}\cdot\bm{e}_{\bm{\Omega}}.
\end{aligned}
}
\label{V_R,s2}
\end{equation}
From Eq.~\eqref{Psi_R_bound} and Cauchy–Schwarz inequality, the lower and upper bounds of $\bm{\mathcal{V}_{\bm{\textit{R},s}}}$ are given by:
\begin{equation}
    \begin{aligned}    
\bm{z}^{\top}_{\bm{\textit{R}}}\bm{\mathfrak{M}}_{\bm{\textit{R}}1}\,\bm{z}_{\bm{\textit{R}}}\leq\bm{\mathcal{V}_{\bm{\textit{R},s}}}\leq&\bm{z}^{\top}_{\bm{\textit{R}}}\bm{\mathfrak{M}}_{\bm{\textit{R}}2}\,\bm{z}_{\bm{\textit{R}}},
    \end{aligned}
    \label{V_R,s_quadratic1}
\end{equation}
where $\bm{z}_{\bm{\textit{R}}}=\left(\|\bm{e}_{\bm{R}}\|,\|\bm{e}_{\bm{\Omega}}\|\right)^{\top}\!\!\in\mathbb{R}^{2}$ and
\begin{equation}
    \begin{aligned}    
      \bm{\mathfrak{M}}_{\bm{\textit{R}}1}={\renewcommand{\arraycolsep}{5pt}\renewcommand{\arraystretch}{1.5}\begin{bmatrix}
\frac{k_{R}}{2}&-\frac{c_R}{2}\\
       -\frac{c_R}{2} &\frac{1}{2}\\
    \end{bmatrix}}
    \end{aligned},
    \begin{aligned}    
      \bm{\mathfrak{M}}_{\bm{\textit{R}}2}={\renewcommand{\arraycolsep}{5pt}\renewcommand{\arraystretch}{1.5}\begin{bmatrix}
\frac{k_{R}}{2-\psi_{\textit{R}}}&\frac{c_R}{2}\\
       \frac{c_R}{2} &\frac{1}{2}\\
    \end{bmatrix}}
    \end{aligned}.
    \label{Matrix_R_1and2}
\end{equation}
If positive constant $c_R$ is chosen sufficiently small to satisfy:
\begin{equation}
{\small
    \begin{aligned}
    c_R\!< \min\left\{\sqrt{k_R},\sqrt{\frac{2k_{R}}{2-\psi_{\textit{R}}}}\right\},
\end{aligned}
}\label{cR_condition}
\end{equation}
matrices $ \bm{\mathfrak{M}}_{\bm{\textit{R}}1}$ and $ \bm{\mathfrak{M}}_{\bm{\textit{R}}2}$ become positive definite, which implies $\bm{\mathcal{V}_{\bm{\textit{R},s}}}$ is positive definite and bounded by:
\begin{equation}
    \begin{aligned}    
\lambda_{\min}( \bm{\mathfrak{M}}_{\bm{\textit{R}}1})\|\bm{z}_{\bm{\textit{R}}}\|^2\leq\bm{\mathcal{V}_{\bm{\textit{R},s}}}\leq\lambda_{\max}(\bm{\mathfrak{M}}_{\bm{\textit{R}}2})\|\bm{z}_{\bm{\textit{R}}}\|^2,
    \end{aligned}
    \label{V_R,s_quadratic2}
\end{equation}
where the $\lambda_{\min}(\bullet)$ and $\lambda_{\max}(\bullet)$ denote the minimum and maximum eigenvalue of a matrix.
\subsubsection{Candidate for Full-State Error }
Further define the  Lyapunov candidate function for the full-state error as:
\begin{equation}
{\small
\begin{aligned}
\bm{\mathcal{V}_{\bm{s}}}=k_{R}\Psi_{\textit{R}}+\bm{\sum}_{j=1}^3\Big{(}\frac{1}{2}\bm{\mathcal{E}}_{\bm{\textit{x}} j}^{\top}\bm{P}_j\bm{\mathcal{E}}_{\bm{\textit{x}} j}+\frac{1}{2}\|\bm{e}^{[j]}_{\bm{\Omega}}\|^2+c_R\bm{e}^{[j]}_{\bm{R}}\bm{e}^{[j]}_{\bm{\Omega}}\Big{)}.
\end{aligned}
}
\label{V_s}
\end{equation}
The $\bm{P}_j\!\in\!\mathbb{R}^{2\times2}$ denotes the $j^{th}$ symmetric positive-definite matrix that satisfies the Lyapunov equation:
\begin{equation}
    \bm{\Lambda}_{\textit{x}j}^{\top}\bm{P}_j
+\bm{P}_j\bm{\Lambda}_{\textit{x}j}=-\bm{Q}_j,
\end{equation}
with the $j^{th}$ positive-definite matrix $\bm{Q}_j>0$.
The compact form of Eq.~\eqref{V_s} can be expressed as:
\begin{equation}
{\small
\begin{aligned}
\bm{\mathcal{V}_{\bm{s}}}=k_{R}\Psi_{\textit{R}}+\frac{1}{2}\|\bm{e}_{\bm{\Omega}}\|^2+c_R\bm{e}_{\bm{R}}\cdot\bm{e}_{\bm{\Omega}}+\bm{\sum}_{j=1}^3\frac{1}{2}\bm{\mathcal{E}}_{\bm{\textit{x}} j}^{\top}\bm{P}_j\bm{\mathcal{E}}_{\bm{\textit{x}} j}.
\end{aligned}
}
\label{V_s2}
\end{equation}
From Eq.~\eqref{Psi_R} and Cauchy–Schwarz inequality, the lower and upper bounds of $\bm{\mathcal{V}_{\bm{s}}}$ are given by:
\begin{equation}
    \begin{aligned}    
\bm{z}^{\top}\bm{\mathfrak{M}}_{1}\bm{z}\leq\bm{\mathcal{V}_{s}}\leq&\bm{z}^{\top}\bm{\mathfrak{M}}_{2}\bm{z},
    \end{aligned}
    \label{V_s_quadratic1}
\end{equation}
where $\bm{z}=\left(\|\bm{\mathcal{E}}_{\bm{\textit{x}} 1}\|,\|\bm{\mathcal{E}}_{\bm{\textit{x}} 2}\|,\|\bm{\mathcal{E}}_{\bm{\textit{x}} 3}\|,\|\bm{e}_{\bm{R}}\|,\|\bm{e}_{\bm{\Omega}}\|\right)^{\top}\!\!\!\in\mathbb{R}^{5}$ and 
\begin{equation}
    \begin{aligned}    
      \bm{\mathfrak{M}}_{1}={\renewcommand{\arraycolsep}{2pt}\begin{bmatrix}
\bm{\mathfrak{M}}_{\bm{\textit{x}}1}&0\\
\\[-2pt]
     0 & \bm{\mathfrak{M}}_{\bm{\textit{R}}1}\\
    \end{bmatrix}}
    \end{aligned},
 \begin{aligned}    
      \bm{\mathfrak{M}}_{2}={\renewcommand{\arraycolsep}{2pt}\begin{bmatrix}
\bm{\mathfrak{M}}_{\bm{\textit{x}}2}&0\\
\\[-2pt]
     0 & \bm{\mathfrak{M}}_{\bm{\textit{R}}2}\\
    \end{bmatrix}}
    \end{aligned},
\end{equation}
where submatrices $ \bm{\mathfrak{M}}_{\bm{\textit{R}}1}$ and $ \bm{\mathfrak{M}}_{\bm{\textit{R}}2}$ have been given in Eq.~\eqref{Matrix_R_1and2} and submatrices $\bm{\mathfrak{M}}_{\bm{\textit{x}}1},\bm{\mathfrak{M}}_{\bm{\textit{x}}2}\in\mathbb{R}^{3\times3}$ are expressed as:
\begin{equation}
    \begin{aligned}    
      \bm{\mathfrak{M}}_{\bm{\textit{x}}1}={\renewcommand{\arraycolsep}{0pt}\begin{bmatrix}
\lambda_{\min}(\bm{P}_1)&0&0\\
0&\lambda_{\min}(\bm{P}_2)&0\\
0&0&\lambda_{\min}(\bm{P}_3)
    \end{bmatrix}},\\
     \bm{\mathfrak{M}}_{\bm{\textit{x}}2}={\renewcommand{\arraycolsep}{0pt}\begin{bmatrix}
\lambda_{\max}(\bm{P}_1)&0&0\\
0&\lambda_{\max}(\bm{P}_2)&0\\
0&0&\lambda_{\max}(\bm{P}_3)
    \end{bmatrix}},
    \end{aligned}
\end{equation}
where, if positive constant $c_R$ satisfies Eq.~\eqref{cR_condition}, all submatrices of $\bm{\mathfrak{M}}_{1}$ and $ \bm{\mathfrak{M}}_{2}$ become positive definite. This implies that $\bm{\mathcal{V}_{s}}$ is positive definite and bounded by:
\begin{equation}
    \begin{aligned}    
\lambda_{\min}( \bm{\mathfrak{M}}_{1})\|\bm{z}\|^2\leq\bm{\mathcal{V}_{\bm{s}}}\leq\lambda_{\max}(\bm{\mathfrak{M}}_{2})\|\bm{z}\|^2.
    \end{aligned}
    \label{V_s_quadratic2}
\end{equation}

\subsubsection{Candidate for Estimation Errors}
Next, we define the Lyapunov candidate function for all estimation errors, which consists of translational and rotational components $\bm{\mathcal{V}}_{\bm{\textit{x}},e}$, $\bm{\mathcal{V}}_{\bm{\textit{R}},e}$:
\begin{equation}
{\footnotesize
\begin{aligned}
\bm{\mathcal{V}_{\bm{e}}}\!=\!\underbrace{\bm{\sum}_{j=1}^3\Big{(}\frac{1}{2}\eta_{\textit{m}}\widetilde{m}_j^2\!+\!\frac{1}{2\gamma_{\bm{\textit{x}}j}}\bm{\tilde{\mathcal{W}}}_{\bm{\textit{x}} j}^{\top}\bm{\tilde{\mathcal{W}}}_{\bm{\textit{x}} j}\Big{)}\!}_{\bm{\mathcal{V}}_{\bm{\textit{x},e}}}+\underbrace{\!\bm{\sum}_{j=1}^3\Big{(}\frac{1}{2}\eta_{J}\widetilde{J}_j^2\!+\!\frac{1}{2\gamma_{\bm{\textit{R}}j}}\bm{\tilde{\mathcal{W}}}_{\bm{\textit{R}} j}^{\top}\bm{\tilde{\mathcal{W}}}_{\bm{\textit{R}} j}\Big{)}}_{\bm{\mathcal{V}}_{\bm{\textit{R},e}}},\\[-10pt]
\end{aligned}
}
\label{V_e}
\end{equation}
where the $\eta_{\textit{m}}$, $\eta_{J}$, $\gamma_{\bm{\textit{x}}j}$ and $\gamma_{\bm{\textit{R}}j}$ are positive constants. 

\subsubsection{Complete Candidate}
Combining Eq.~\eqref{V_s} and Eq.~\eqref{V_e}, the Lyapunov candidate function for complete dynamics is rearranged and given as follows: 
\begin{equation}
{\small
\begin{aligned}
\bm{\mathcal{V}}=&\underbrace{\bm{\sum}_{j=1}^3\Big{(}\frac{1}{2}\bm{\mathcal{E}}_{\bm{\textit{x}} j}^{\top}\bm{P}_j\bm{\mathcal{E}}_{\bm{\textit{x}} j}+\frac{1}{2}\eta_{\textit{m}}\widetilde{m}_j^2+ \frac{1}{2\gamma_{\bm{\textit{x}}j}}\bm{\tilde{\mathcal{W}}}_{\bm{\textit{x}} j}^{\top}\bm{\tilde{\mathcal{W}}}_{\bm{\textit{x}} j}\Big{)}}_{\text{Translational candidate function :}\bm{\mathcal{V}}_{\bm{\textit{x}}}}\\
&\underbrace{\!\!\!+k_{R}\Psi_{\textit{R}}\!+\!\!\bm{\sum}_{j=1}^3\!\!\Big{(}\frac{1}{2}\|\bm{e}^{[j]}_{\bm{\Omega}}\|^2\!\!+\!c_R\bm{e}^{[j]}_{\bm{R}}\bm{e}^{[j]}_{\bm{\Omega}}\!\!+\!\frac{1}{2}\eta_{J}\widetilde{J}_j^2\!+\!\frac{1}{2\gamma_{\bm{\textit{R}}j}}\bm{\tilde{\mathcal{W}}}_{\bm{\textit{R}} j}^{\top}\bm{\tilde{\mathcal{W}}}_{\bm{\textit{R}} j}\!\Big{)}}_{\text{Rotational candidate function:}\bm{\mathcal{V}}_{\bm{\textit{R}}}},
\end{aligned}
}
\label{V}
\end{equation}
where the first line denotes the translational candidate function $\bm{\mathcal{V}}_{\bm{\textit{x}}}$, while the second line corresponds to the rotational candidate function $\bm{\mathcal{V}}_{\bm{\textit{R}}}$. 

From Eqs.~\eqref{V_R,s}, \eqref{V_R,s_quadratic2}, \eqref{V_e} and \eqref{V}, it holds that:
\begin{equation}
{\small
\begin{aligned}
     \lambda_{\min}( \bm{\mathfrak{M}}_{\bm{\textit{R}}1})\|\bm{z}_{\bm{\textit{R}}}\|^2\!+\!\bm{\mathcal{V}}_{\bm{\textit{R},e}}\leq\bm{\mathcal{V}}_{\bm{\textit{R}}}\leq\lambda_{\max}(\bm{\mathfrak{M}}_{\bm{\textit{R}}2})\|\bm{z}_{\bm{\textit{R}}}\|^2\!+\!\bm{\mathcal{V}}_{\bm{\textit{R},e}}.
\end{aligned}
}
\label{V_R_bound1}
\end{equation}

From Eqs.~\eqref{V_s}, \eqref{V_s_quadratic2}, \eqref{V_e} and \eqref{V}, it holds that:
\begin{equation}
{\small
\begin{aligned}
     \lambda_{\min}( \bm{\mathfrak{M}}_{1})\|\bm{z}\|^2\!+\!\bm{\mathcal{V}_{\bm{e}}}\leq\bm{\mathcal{V}}\leq\lambda_{\max}(\bm{\mathfrak{M}}_{2})\|\bm{z}\|^2\!+\!\bm{\mathcal{V}_{\bm{e}}}.
\end{aligned}
}
\label{V_bound1}
\end{equation}
\textbf{\textit{Lemma 1:}} Given that $\bm{\mathcal{V}}_{\bm{\textit{R},e}}$ and  $\bm{\mathcal{V}}_{\bm{e}}$ are positive-definite and bounded, it holds that there always exist positive constants $\mathfrak{p}_1$, $\mathfrak{p}_2$, $\mathfrak{p}_3$ and $\mathfrak{p}_4$ outside the ball with arbitrary bounded radius $\epsilon$ around $\bm{z}_{\bm{\textit{R}}}=0$ and $\bm{z}=0$  such that:
\begin{equation}
{\small
\begin{aligned}
     \mathfrak{p}_1\lambda_{\min}( \bm{\mathfrak{M}}_{\bm{\textit{R}}1})\|\bm{z}_{\bm{\textit{R}}}\|^2\!\leq\bm{\mathcal{V}}_{\bm{\textit{R}}}\leq\mathfrak{p}_2\lambda_{\max}(\bm{\mathfrak{M}}_{\bm{\textit{R}}2})\|\bm{z}_{\bm{\textit{R}}}\|^2,
\end{aligned}
}
\label{VR_p1p2bound}
\end{equation}
\begin{equation}
{\small
\begin{aligned}
     \mathfrak{p}_3\lambda_{\min}( \bm{\mathfrak{M}}_{1})\|\bm{z}\|^2\!\leq\bm{\mathcal{V}}\leq\mathfrak{p}_4\lambda_{\max}(\bm{\mathfrak{M}}_{2})\|\bm{z}\|^2.
\end{aligned}
}
\label{V_p1p2bound}
\end{equation}
\textbf{\textit{Proof of Lemma 1:}}  With the fact that $\bm{\mathcal{V}}_{\bm{\textit{R},e}}$ and  $\bm{\mathcal{V}}_{\bm{e}}$ are positive-definite and designed to be bounded, if $\|\bm{z}_{\bm{\textit{R}}}\|\geq\epsilon$, $\|\bm{z}\|\geq\epsilon$ and radius $\epsilon$ is positive and bounded, there always exist sufficiently large but bounded positive constants $\mathfrak{p}_1$, $\mathfrak{p}_2$, $\mathfrak{p}_3$ and $\mathfrak{p}_4$ to satisfy:
\begin{equation}
{\small
\begin{aligned}
     \mathfrak{p}_1&\leq1+\frac{\bm{\mathcal{V}}_{\bm{\textit{R},e}}}{\lambda_{\min}( \bm{\mathfrak{M}}_{\bm{\textit{R}}1})\|\bm{z}_{\bm{\textit{R}}}\|^2}, \\
     \mathfrak{p}_2&\geq1+\frac{\bm{\mathcal{V}}_{\bm{\textit{R},e}}}{\lambda_{\max}( \bm{\mathfrak{M}}_{\bm{\textit{R}}2})\|\epsilon\|^2}\geq1+\frac{\bm{\mathcal{V}}_{\bm{\textit{R},e}}}{\lambda_{\max}( \bm{\mathfrak{M}}_{\bm{\textit{R}}2})\|\bm{z}_{\bm{\textit{R}}}\|^2},\\
     \mathfrak{p}_3&\leq1+\frac{\bm{\mathcal{V}}_{\bm{e}}}{\lambda_{\min}( \bm{\mathfrak{M}}_{1})\|\bm{z}\|^2}, \\
     \mathfrak{p}_4&\geq1+\frac{\bm{\mathcal{V}}_{\bm{e}}}{\lambda_{\max}( \bm{\mathfrak{M}}_{2})\|\epsilon\|^2}\geq1+\frac{\bm{\mathcal{V}}_{\bm{e}}}{\lambda_{\max}( \bm{\mathfrak{M}}_{2})\|\bm{z}\|^2}.\\
\end{aligned}
}
\label{Proof_of_Corollary1}
\end{equation}
Substituting these to Eqs.~\eqref{V_R_bound1} and \eqref{V_bound1} yields Eqs.~\eqref{VR_p1p2bound} and \eqref{V_p1p2bound}. Therefore, \textbf{\textit{Lemma 1}} is established.

\textbf{\textit{Remark 2:}} We can choose smaller constants $\eta_{\textit{m}}$, $\eta_{J}$, $1/\gamma_{\bm{\textit{x}}j}$ and $1/\gamma_{\bm{\textit{R}}j}$ to obtain smaller $\mathfrak{p}_1$, $\mathfrak{p}_2$, $\mathfrak{p}_3$ and $\mathfrak{p}_4$.  

\subsection{Stability Proof}
\subsubsection{\textbf{Proof of Proposition 1}}
The time-derivative of the rotational candidate function is driven with the fact that $\dot{\Psi}_{\textit{R}}=\bm{e}_{\bm{R}}\cdot\bm{e}_{\bm{\Omega}}$ \cite{2010 Geometric tracking control of a quadrotor UAV on SE(3)}:
\begin{equation}
    {\small
    \begin{aligned}    \bm{\dot{\mathcal{V}}}_{\bm{\textit{R}}}\!=&k_{R}\bm{e}_{\bm{R}}\cdot\!\bm{e}_{\bm{\Omega}}+\bm{\sum}_{j=1}^3\Bigg{\{}\bm{e}_{\bm{\Omega}}^{[j]}\bm{\dot{e}}_{\bm{\Omega}}^{[j]}+c_R\bm{e}^{[j]}_{\bm{\Omega}}\bm{\dot{e}}^{[j]}_{\bm{R}}+c_R\bm{e}^{[j]}_{\bm{R}}\bm{\dot{e}}^{[j]}_{\bm{\Omega}}\\
&+\eta_{J}\widetilde{J}_j\frac{\bm{\dot{\bar{\mathit{J}}}}^{[j]}}{\bm{\bar{J}}^{[j]^2}}-\frac{1}{\gamma_{\bm{\textit{R}}j}}\big{(}\bm{\mathcal{W}}^*_{\bm{\textit{R}} j}\!-\!\bm{\bar{\mathcal{W}}}_{\bm{\textit{R}} j}\big{)}\!^{\top}\bm{\dot{\bar{\mathcal{W}}}}_{\bm{\textit{R}} j}
    \Bigg{\}}\\
    =&\bm{\sum}_{j=1}^3\Bigg{\{}k_{R}\bm{e}_{\bm{R}}^{[j]}\bm{e}_{\bm{\Omega}}^{[j]}+(\bm{e}_{\bm{\Omega}}^{[j]}+c_R\bm{e}^{[j]}_{\bm{R}})\bm{\dot{e}}_{\bm{\Omega}}^{[j]}+c_R\bm{e}^{[j]}_{\bm{\Omega}}\bm{\dot{e}}^{[j]}_{\bm{R}}\\
    &+\eta_{J}\widetilde{J}_j\frac{\bm{\dot{\bar{\mathit{J}}}}^{[j]}}{\bm{\bar{J}}^{[j]^2}}-\frac{1}{\gamma_{\bm{\textit{R}}j}}\big{(}\bm{\mathcal{W}}^*_{\bm{\textit{R}} j}-\bm{\bar{\mathcal{W}}}_{\bm{\textit{R}} j}\big{)}^{\top}\bm{\dot{\bar{\mathcal{W}}}}_{\bm{\textit{R}} j}
    \Bigg{\}}.
    \end{aligned}
    }   
\end{equation}
Substituting Eq.~\eqref{rotational error dynamics5}, we derive:
\begin{equation}
    {\small
    \begin{aligned}    \bm{\dot{\mathcal{V}}}_{\bm{\textit{R}}}\!=&\bm{\sum}_{j=1}^3\!\Bigg{\{}\!\!-k_{R}c_{R}\|\bm{e}_{\bm{R}}^{[j]}\|^{2}\!-\!k_{\Omega}\|\bm{e}_{\bm{\Omega}}^{[j]}\|^{2}\!-\!k_{\Omega}c_{R}\bm{e}_{\bm{\Omega}}^{[j]}\bm{e}_{\bm{R}}^{[j]}+c_R\bm{e}_{\bm{\Omega}}^{[j]}\bm{\dot{e}}_{\bm{R}}^{[j]}\\[-5pt]
&+\!\!\left(\!\bm{e}^{[j]}_{\bm{\Omega}}\!+\!c_R\bm{e}^{[j]}_{\bm{R}}\!\right)\!\!\bigg{\{}\bm{\varpi}^{[j]}_{\bm{\textit{R}}}\!+\!\left(\bm{J}^{-1}\!\!\bm{\Delta_{\mathrm{M}}}\right)^{[j]}\!\!\bigg{\}}\\[-5pt]
    &+\widetilde{J}_j
\bigg{\{}\left(\bm{e}^{[j]}_{\bm{\Omega}}+c_R\bm{e}^{[j]}_{\bm{R}}\right)\bm{\mathrm{M}_d}^{[j]}+\eta_{J}\frac{\bm{\dot{\bar{\mathit{J}}}}^{[j]}}{\bm{\bar{J}}^{[j]^2}}\bigg{\}}\\
    &\!\!\!\!\!\!\!+\frac{1}{\gamma_{\textit{R}j}}\left(\bm{\mathcal{W}}^*_{\bm{\textit{R}} j}-\bm{\bar{\mathcal{W}}}_{\bm{\textit{R}} j}\right)^{\top}\!\bigg{\{}\gamma_{\bm{\textit{R}}j}\left(\bm{e}^{[j]}_{\bm{\Omega}}+c_R\bm{e}^{[j]}_{\bm{R}}\right)\bm{\hbar}(\textbf{x}_{\textit{R}j})\!-\!\bm{\dot{\bar{\mathcal{W}}}}_{\textit{R}j}\bigg{\}}\!\Bigg{\}}\\
    &+(\bm{e}_{\bm{\Omega}}+c_R\bm{e}_{\bm{R}})(\underbrace{\bm{J}^{-1}[\bm{\Omega}]_{\times}\bm{J}\bm{\Omega}}_{\text{if $\bm{J}$ is known}}-\bm{J}^{-1}[\bm{\Omega}]_{\times}\bm{J}\bm{\Omega}).\\[-10pt]
    \end{aligned}
    }
    \label{dot_V_R}
\end{equation}
In both \textbf{\textit{Scenario 1 ($\bm{J}$ is known)}} and \textbf{\textit{Scenario 2 ($\bm{J}$ is unknown)}}, the last line of Eq.~\eqref{dot_V_R} can be canceled (see Appendix. \ref{Rotational Error Dynamics}).
  In addition, since $\bm{\dot{\bar{\mathit{J}}}}^{[j]}$ and $\bm{\dot{\bar{\mathcal{W}}}}_{\bm{\textit{R}} 
 j}$ are designed as in Eqs.~\eqref{Adaptive Law of Inertia Tensor} and \eqref{Estimated Weights_R}, the last three lines vanish.
 Given that $\|\bm{\Delta_{\mathrm{M}}}\|$ converges to zero if the reference coefficients $c'_M$ is optimally chosen, we consider the lower and upper bounds of  $\left(\bm{J}^{-1}\!\bm{\Delta_{\mathrm{M}}}\right)\!^{[j]}$ as follows:
 \begin{equation}
{\small
\begin{aligned}
0\leq\|\left(\bm{J}^{-1}\!\bm{\Delta_{\mathrm{M}}}\right)\!^{[j]}\|\leq\|\frac{\bm{\Delta_{\mathrm{M}}}}{\lambda_{\min}(\bm{J})}\|\leq\frac{\varepsilon_{\mathbf{M}}}{\lambda_{\min}(\bm{J})},
\end{aligned}
}
\label{DeltaM_bound}
\end{equation}
where $\varepsilon_{\mathbf{M}}$ is defined as the upper bound of  $\|\bm{\Delta_{\mathrm{M}}}\|$. 

Then, since $\|\bm{e_R}\|<1$ and $\|\bm{\dot{e}}_{\bm{R}}\|\leq\|\bm{e}_{\bm{\Omega}}\|$ from Eq.~\eqref{rotational error dynamics1}, we can apply the foregoing bounds from Eq.~\eqref{DeltaM_bound} to obtain the upper bound of $\bm{\dot{\mathcal{V}}}_{\bm{\textit{R}}}$:
\begin{equation}
    {\footnotesize
    \begin{aligned}    
    \bm{\dot{\mathcal{V}}}_{\bm{\textit{R}}}\leq
    &\!-\!k_{R}c_{R}\|\bm{e}_{\bm{R}}\|^{2}\!-\!\left(k_{\Omega}\!-\!c_R\right)\|\bm{e}_{\bm{\Omega}}\|^{2}\!+\!k_{\Omega}c_{R}\|\bm{e}_{\bm{\Omega}}\|\|\bm{e}_{\bm{R}}\|\!+\!c_R\|\bm{e}_{\bm{\Omega}}\|^2\\
    &+c_R\bigg{(}\varepsilon_{\bm{\textit{R}}}+\frac{\varepsilon_{\mathbf{M}}}{\lambda_{\min}(\bm{J})}\bigg{)}\|\bm{e}_{\bm{R}}\|+\bigg{(}\varepsilon_{\bm{\textit{R}}}+\frac{\varepsilon_{\mathbf{M}}}{\lambda_{\min}(\bm{J})}\bigg{)}\|\bm{e}_{\bm{\Omega}}\|.
    \end{aligned}
    } 
    \label{dot_V_R_bound1}
\end{equation}
where $\varepsilon_{{\bm{\textit{R}}}}\!\in\!\mathbb{R}$ is defined as the upper bound of optimal approximation error $\bm{\varpi}_{\textit{R}}$ of neural networks:
\begin{equation}
{\small
\begin{aligned}
\|\bm{\varpi}^{[j]}_{\textit{R}}\|\leq\|\bm{\varpi}_{\textit{R}}\|
\leq\varepsilon_{{\bm{\textit{R}}}}.
\end{aligned}
}
\label{varpix_bound}
\end{equation}
By choosing $k_\Omega > c_R$, we can apply Young’s inequality to the last line, yielding:
\begin{equation}
    {\footnotesize
    \begin{aligned}    
&c_R\bigg{(}\varepsilon_{\bm{\textit{R}}}+\frac{\varepsilon_{\mathbf{M}}}{\lambda_{\min}(\bm{J})}\bigg{)}\|\bm{e}_{\bm{R}}\|\leq\frac{c_R^2\bigg{(}\varepsilon_{\bm{\textit{R}}}+\frac{\varepsilon_{\mathbf{M}}}{\lambda_{\min}(\bm{J})}\bigg{)}^2}{2k_{R}c_{R}}+\frac{k_{R}c_{R}}{2}\|\bm{e}_{\bm{R}}\|^{2},\\
    &\bigg{(}\!\varepsilon_{\bm{\textit{R}}}\!+\!\frac{\varepsilon_{\mathbf{M}}}{\lambda_{\min}(\bm{J})}\!\bigg{)}\|\bm{e}_{\bm{\Omega}}\|\!\leq\!\frac{\bigg{(}\varepsilon_{\bm{\textit{R}}}\!+\!\frac{\varepsilon_{\mathbf{M}}}{\lambda_{\min}(\bm{J})}\bigg{)}^2}{2\left(k_{\Omega}\!\!-\!c_R\right)}\!+\!\frac{k_{\Omega}\!-\!c_R\!}{2}\|\bm{e}_{\bm{\Omega}}\|^2.
    \end{aligned}
    } 
    \label{dot_V_R_bound2}
\end{equation}
From here, we can reformulate Eq.~\eqref{dot_V_R_bound1} into:
\begin{equation}
    \begin{aligned}    
    \bm{\dot{\mathcal{V}}}_{\bm{\textit{R}}}\leq&-\bm{z}^{\top}_{\bm{\textit{R}}}\!\bm{\mathcal{M}}_{\bm{\textit{R}}}\,\bm{z}_{\bm{\textit{R}}}+\mathbf{C}_{\bm{\textit{R}}},
    \end{aligned}
    \label{dot_V_R_quadratic}
\end{equation}
where $\bm{z}_{\bm{\textit{R}}}=\left(\|\bm{e}_{\bm{R}}\|,\|\bm{e}_{\bm{\Omega}}\|\right)^{\top}\!\!\in\mathbb{R}^{2}$. The matrix $\bm{\mathcal{M}}_{\bm{\textit{R}}}\in\mathbb{R}^{2\times2}$ is given by:
\begin{equation}
    \begin{aligned}    
       \bm{\mathcal{M}}_{\bm{\textit{R}}}={\renewcommand{\arraycolsep}{5pt}\renewcommand{\arraystretch}{1.5}\begin{bmatrix}
\frac{k_Rc_R}{2}&\frac{-k_{\Omega}c_R}{2}\\
       \frac{-k_{\Omega}c_R}{2} &\frac{k_{\Omega}-c_R}{2}\\
    \end{bmatrix}}
    \end{aligned},
    \label{Matrix_R}
\end{equation}
and the constant term is expressed as:
\begin{equation}
    \begin{aligned}    
\mathbf{C}_{\bm{\textit{R}}}=\frac{c_R\bigg{(}\varepsilon_{\bm{\textit{R}}}+\frac{\varepsilon_{\mathbf{M}}}{\lambda_{\min}(\bm{J})}\bigg{)}^2}{2k_{R}}+\frac{\bigg{(}\varepsilon_{\bm{\textit{R}}}+\frac{\varepsilon_{\mathbf{M}}}{\lambda_{\min}(\bm{J})}\bigg{)}^2}{2\left(k_{\Omega}\!-\!c_R\!\right)}.
    \end{aligned}
    \label{C_R}
\end{equation}
Combining with Eq.~\eqref{cR_condition}, if positive constant $c_R$ is sufficiently small to satisfy:
\begin{equation}
{\small
    \begin{aligned}
    c_R\!< \min\left\{\frac{k_Rk_{\Omega}}{k_{\Omega}^2+k_R},\sqrt{k_R},\sqrt{\frac{2k_{R}}{2-\psi_{\textit{R}}}},k_\Omega\right\},\label{cR bound}
\end{aligned}
}
\end{equation}
it follows that matrix $\bm{\mathcal{M}}_{\bm{\textit{R}}}$ is positive-definite. Therefore, Eq.~\eqref{dot_V_R_quadratic} can be further expressed as:
\begin{equation}
    \begin{aligned}    
    \bm{\dot{\mathcal{V}}}_{\bm{\textit{R}}}\leq-\lambda_{\min}(\bm{\mathcal{M}}_{\bm{\textit{R}}})\|\bm{z}_{\bm{\textit{R}}}\|^2+\mathbf{C}_{\bm{\textit{R}}},
    \end{aligned}
    \label{dot_V_R_quadratic2}
\end{equation}
where $\mathbf{C}_{\bm{\textit{R}}}>0$. To proceed, substituting Eq.~\eqref{VR_p1p2bound}, it holds that:
\begin{equation}
    \begin{aligned}    
    \bm{\dot{\mathcal{V}}}_{\bm{\textit{R}}}\leq-2\beta_1\bm{\mathcal{V}_{\bm{\textit{R}}}}+\mathbf{C}_{\bm{\textit{R}}},
    \end{aligned}
    \label{dot_V_R_quadratic2}
\end{equation}
with $\beta_1=\frac{\lambda_{\min}(\bm{\mathcal{M}}_{\bm{\textit{R}}})}{2\mathfrak{p}_2\lambda_{\max}(\bm{\mathfrak{M}}_{\bm{\textit{R}}2})}$. This drives that:
\begin{equation}
    \begin{aligned}    
    \|\bm{z}_{\bm{\textit{R}}}(t)\|\leq\alpha_1\|\bm{z}_{\bm{\textit{R}}}(0)\|e^{-\beta_1 t}+\epsilon_1,
    \end{aligned}
    \label{NES condition1}
\end{equation}
where $\bm{z}_{\bm{\textit{R}}}(0)\!\!\in\!\mathcal{D}_{\bm{\textit{R}}0}$,   $\alpha_1\!\!=\!\!\sqrt{\frac{\mathfrak{p}_2\lambda_{\max}(\bm{\mathfrak{M}}_{\bm{\textit{R}}2})}{\mathfrak{p}_1\lambda_{\min}(\bm{\mathfrak{M}}_{\bm{\textit{R}}1})}}$, $\epsilon_1\!\!=\!\!\sqrt{\frac{\mathbf{C}_{\bm{\textit{R}}}}{2\beta_1\mathfrak{p}_1\lambda_{\min}(\bm{\mathfrak{M}}_{\bm{\textit{R}}1})}}$. 

By properly selecting the parameters of the neural network, including the number of neurons $l$ in the hidden layer, the center vectors $\textbf{c}_{k}$ and the width $b_k$, the universal approximation theorem \cite{1989 Multilayer feedforward networks are universal approximators} ensures that the upper bound of the approximation error can be reduced arbitrarily small, i.e., $\varepsilon_{\bm{\textit{R}}}\to0^+$. Furthermore, precise identification of the reference coefficient $c'_M$ through aerodynamic experiments drives the upper bound $\varepsilon_{\mathbf{M}}$ to converge to zero. 
As $\varepsilon_{\bm{\textit{R}}}\to0^+$, $\varepsilon_{\mathbf{M}}\to0$, it follows that $\mathbf{C}_{\bm{\textit{R}}}\to0^+$ and $\epsilon_1\to0^+$. Therefore, \textbf{\textit{Proposition 1}} is established. 

\subsubsection{\textbf{Proof of Propositions 2 and 0}}
The time-derivative of the translational candidate function is given by:
\begin{equation}
    {\small
    \begin{aligned}    \bm{\dot{\mathcal{V}}}_{\bm{\textit{x}}}\!=&\!\bm{\sum}_{j=1}^3\Bigg{\{}\frac{1}{2}\bm{\dot{\mathcal{E}}}_{\bm{\textit{x}} j}^{\top}\bm{P}_j\bm{\mathcal{E}}_{\bm{\textit{x}} j}\!+\!\frac{1}{2}\bm{\mathcal{E}}_{\bm{\textit{x}} j}^{\top}\bm{P}_j\bm{\dot{\mathcal{E}}}_{\bm{\textit{x}} j}
   +\eta_{\textit{m}}\widetilde{m}_j\frac{\bm{\dot{\bar{m}}}^{[j]}}{\bm{\bar{m}}^{[j]^2}}\\
    &-\frac{1}{\gamma_{\bm{\textit{x}}j}}\big{(}\bm{\mathcal{W}}^*_{\bm{\textit{x}} j}\!-\!\bm{\bar{\mathcal{W}}}_{\bm{\textit{x}} j}\big{)}\!^{\top}\bm{\dot{\bar{\mathcal{W}}}}_{\bm{\textit{x}} j}
    \Bigg{\}}.   
    \end{aligned}
    }   
\end{equation}
Substituting Eq.~\eqref{translational error dynamics3}, we obtain:
\begin{equation}
    {\small
    \begin{aligned}    \bm{\dot{\mathcal{V}}}_{\bm{\textit{x}}}\!=&\!\bm{\sum}_{j=1}^3\Bigg{\{}\frac{1}{2}\left(\bm{{\mathcal{E}}}_{\bm{\textit{x}} j}^{\top}\bm{\Lambda}_{\bm{\textit{x}} j}^{\top}+G_j^{\top}\right)\bm{P}_j\bm{\mathcal{E}}_{\bm{\textit{x}} j}\!+\!\frac{1}{2}\bm{\mathcal{E}}_{\bm{\textit{x}} j}^{\top}\bm{P}_j\left(\bm{\Lambda}_{\bm{\textit{x}} j}\bm{\mathcal{E}}_{\bm{\textit{x}} j}+G_j\right)\\
   &+\eta_{\textit{m}}\widetilde{m}_j\frac{\bm{\dot{\bar{m}}}^{[j]}}{\bm{\bar{m}}^{[j]^2}}
    -\frac{1}{\gamma_{\bm{\textit{x}}j}}\big{(}\bm{\mathcal{W}}^*_{\bm{\textit{x}} j}\!-\!\bm{\bar{\mathcal{W}}}_{\bm{\textit{x}} j}\big{)}\!^{\top}\bm{\dot{\bar{\mathcal{W}}}}_{\bm{\textit{x}} j}
    \Bigg{\}}\\
    =&\!\bm{\sum}_{j=1}^3\Bigg{\{}\frac{1}{2}\bm{{\mathcal{E}}}_{\bm{\textit{x}} j}^{\top}\left(\bm{\Lambda}_{\bm{\textit{x}} j}^{\top}\bm{P}_j+\bm{P}_j\bm{\Lambda}_{\bm{\textit{x}} j}\right)\bm{\mathcal{E}}_{\bm{\textit{x}} j}\!+\!\frac{1}{2}G_j^{\top}\bm{P}_j\bm{{\mathcal{E}}}_{\bm{\textit{x}} j}\!+\!\frac{1}{2}\bm{{\mathcal{E}}}_{\bm{\textit{x}} j}^{\top}\bm{P}_jG_j\\
   &+\eta_{\textit{m}}\widetilde{m}_j\frac{\bm{\dot{\bar{m}}}^{[j]}}{\bm{\bar{m}}^{[j]^2}}
    -\frac{1}{\gamma_{\bm{\textit{x}}j}}\big{(}\bm{\mathcal{W}}^*_{\bm{\textit{x}} j}\!-\!\bm{\bar{\mathcal{W}}}_{\bm{\textit{x}} j}\big{)}\!^{\top}\bm{\dot{\bar{\mathcal{W}}}}_{\bm{\textit{x}} j}
    \Bigg{\}}\\
    =&\!\bm{\sum}_{j=1}^3\Bigg{\{}-\frac{1}{2}\bm{{\mathcal{E}}}_{\bm{\textit{x}} j}^{\top}\bm{Q}_j\bm{\mathcal{E}}_{\bm{\textit{x}} j}+\bm{{\mathcal{E}}}_{\bm{\textit{x}} j}^{\top}\bm{P}_jG_j\\
   &+\eta_{\textit{m}}\widetilde{m}_j\frac{\bm{\dot{\bar{m}}}^{[j]}}{\bm{\bar{m}}^{[j]^2}}
    -\frac{1}{\gamma_{\bm{\textit{x}}j}}\big{(}\bm{\mathcal{W}}^*_{\bm{\textit{x}} j}\!-\!\bm{\bar{\mathcal{W}}}_{\bm{\textit{x}} j}\big{)}\!^{\top}\bm{\dot{\bar{\mathcal{W}}}}_{\bm{\textit{x}} j}
    \Bigg{\}}.\\
    \end{aligned}
    }   
\end{equation}
Replacing $G_j$ with its original value from Eq.~\eqref{translational error dynamics2}, we obtain:
\begin{equation}
    {\small
    \begin{aligned}    \bm{\dot{\mathcal{V}}}_{\bm{\textit{x}}}\!=&\!\bm{\sum}_{j=1}^3\Bigg{\{}-\frac{1}{2}\bm{\mathcal{E}}_{\bm{\textit{x}} j}^{\top}\bm{Q}_j\bm{\mathcal{E}}_{\bm{\textit{x}} j}
      +\widetilde{m}_j\left(\bm{\mathcal{E}}_{\bm{\textit{x}} j}^{\top}\bm{P}_j\bm{B} \bm{\mathrm{F}_d}^{[j]}+\eta_{\textit{m}}\frac{\bm{\dot{\bar{m}}}^{[j]}}{\bm{\bar{m}}^{[j]^2}}\right)\\
       &+\frac{1}{\gamma_{\textit{x}j}}\left(\bm{\mathcal{W}}^*_{\bm{\textit{x}} j}-\bm{\bar{\mathcal{W}}}_{\bm{\textit{x}} j}\right)^{\top}\left(\gamma_{\textit{x}j}\bm{\mathcal{E}}_{\bm{\textit{x}} j}^{\top}\bm{P}_j\bm{B}\bm{\hbar}(\textbf{x}_{\textit{x}j})-\bm{\dot{\bar{\mathcal{W}}}}_{\textit{x}j}\right)\\
       &+\bm{\mathcal{E}}_{\bm{\textit{x}} j}^{\top}\bm{P}_j\bm{B}\bigg{\{}\bm{\varpi}^{[j]}_{\textit{x}}-\frac{1}{m}\left(\Delta_f \bm{R}\bm{\vec{e}}_3\right)^{[j]}-\frac{1}{m}\mathcal{X}^{[j]}\bigg{\}}\Bigg{\}},
    \end{aligned}
    }  
    \label{dot_V_x}
\end{equation}
where, since $\bm{\dot{\bar{m}}}^{[j]}$ and $\bm{\dot{\bar{\mathcal{W}}}}_{\bm{\textit{x}} 
 j}$ are designed as in Eqs.~\eqref{Adaptive Law of mass} and \eqref{Estimated Weights_x}, the second and third terms vanish. 
 The upper bound of the $\mathcal{X}$ is expressed as:
\begin{equation}
{\small
\begin{aligned}
\|\mathcal{X}\|=&\|\bm{\mathrm{F}_d}\|\|\left(\bm{\vec{e}}_3^{\top}\bm{R}_{\bm{c}}^{\top}\bm{R}\bm{\vec{e}}_3\right)\bm{R}\bm{\vec{e}}_3-\bm{R}_{\bm{c}}\bm{\vec{e}}_3\|
\leq\|\bm{\mathrm{F}_d}\|\|\bm{e_R}\|
\end{aligned}
}
\label{X_bound1}
\end{equation}
Since $\|\mathcal{X}\|\leq\|\mathcal{X}\|_{1}\leq\sqrt{3}\|\mathcal{X}\|$,  $\|\bm{e_R}\|\leq\|\bm{e_R}\|_{1}\leq\sqrt{3}\|\bm{e_R}\|$ and $\|\bm{\mathrm{F}_d}\|\leq\|\bm{\mathrm{F}_d}\|_{1}$, it holds that:
\begin{equation}
{\small
\begin{aligned}
\|\mathcal{X}^{[j]}\|\leq\|\mathcal{X}\|_{1}
\leq\sqrt{3}\|\bm{\mathrm{F}_d}\|\|\bm{e_R}\|\leq\sqrt{3}\|\bm{\mathrm{F}_d}\|_{1}\cdot\|\bm{e_R}\|.
\end{aligned}
}
\label{X_bound2}
\end{equation}
Now, we consider the upper bound of $\mathcal{X}^{[j]}$:
\begin{equation}
{\small
\begin{aligned}
\|\mathcal{X}^{[j]}\|&\leq
\bm{\sum}_{j=1}^3\sqrt{3}\|\bm{\mathrm{F}_d}^{[j]}\|\|\bm{e_R}\|\\
&\!\!\!\!\!\!\!\!\!\!\!\!\!\!\leq\bm{\sum}_{j=1}^3\sqrt{3\|\bm{\bar{m}}^{[j]}\|^2\left(-\bm{\mathcal{K}}_{\bm{\textit{x}} j}^{\top}\bm{\mathcal{E}}_{\bm{\textit{x}} j}+\bm{\ddot{x}}_{{\bm{d}}}^{[j]}-g\bm{\delta}_{j3} -\bm{\bar{\phi}}_{\bm{\textit{x}}}^{[j]}\right)^2}\cdot\|\bm{e_R}\|\\
&\leq\bm{\sum}_{j=1}^3\sqrt{3}\,\overset{\tiny \text{max}}{m}\|-\bm{\mathcal{K}}_{\bm{\textit{x}} j}^{\top}\bm{\mathcal{E}}_{\bm{\textit{x}} j}+\bm{\ddot{x}}_{{\bm{d}}}^{[j]}-g\bm{\delta}_{j3} -\bm{\bar{\phi}}_{\bm{\textit{x}}}^{[j]}\|\|\bm{e_R}\|\\
&\leq\bm{\sum}_{j=1}^3\sqrt{3}\,\overset{\tiny \text{max}}{m}\left(\|\bm{\mathcal{K}}_{\bm{\textit{x}} j}\|\|\bm{\mathcal{E}}_{\bm{\textit{x}} j}\|+\varepsilon_{c}\right)\|\bm{e_R}\|\\
&\leq\sqrt{3}\,\overset{\tiny \text{max}}{m}\left(\varepsilon_{u}+\varepsilon_{c}\right)\|\bm{e_R}\|\\
\end{aligned}
}
\label{Xj_bound}
\end{equation}
where
\begin{equation}
{\small
\begin{aligned}
\overset{\tiny \text{min}}{m}\leq\|\bm{\bar{m}}^{[j]}\|\leq\overset{\tiny \text{max}}{m},
\end{aligned}
}
\label{m_bar_bound}
\end{equation}
and $\varepsilon_{u}$, $\varepsilon_{c}\in\mathbb{R}$ are defined as the upper bounds of PD control input term and compensation term, respectively:
\begin{equation}
{\small
\begin{aligned}
\bm{\sum}_{j=1}^3\|\bm{\mathcal{K}}_{\bm{\textit{x}} j}\|\|\bm{\mathcal{E}}_{\bm{\textit{x}} j}\|\leq\varepsilon_{u},
\end{aligned}
}
\label{varepsilon_u_bound}
\end{equation}
\begin{equation}
{\small
\begin{aligned}
\bm{\sum}_{j=1}^3\|\bm{\ddot{x}}_{{\bm{d}}}^{[j]}-g\bm{\delta}_{j3} -\bm{\bar{\phi}}_{\bm{\textit{x}}}^{[j]}\|
\leq\varepsilon_{c}.
\end{aligned}
}
\label{varepsilon_c_bound}
\end{equation}

Next, since $\|\Delta_f\|$ converge to zero if the reference coefficients $c'_T$ are optimally chosen, we consider the lower and upper bound of $\left(\Delta_f \bm{R}\bm{\vec{e}}_3\right)\!^{[j]}$ as follows:
\begin{equation}
{\small
\begin{aligned}
0\leq\|\left(\Delta_f \bm{R}\bm{\vec{e}}_3\right)\!^{[j]}\|\leq\|\Delta_f\|\leq\varepsilon_{f},
\end{aligned}
}
\label{Deltaf_bound}
\end{equation}
where $\varepsilon_{f}\in\mathbb{R}$ is defined as the upper bound of  $\|\Delta_f\|$.

Then, since $\|\bm{e_R}\|<1$ and $\|\bm{\dot{e}}_{\bm{R}}\|\leq\|\bm{e}_{\bm{\Omega}}\|$ from Eq.~\eqref{rotational error dynamics1}, we can apply the foregoing bounds  \eqref{dot_V_R_bound1},  \eqref{Xj_bound} and \eqref{Deltaf_bound} to obtain the upper bound of $\bm{\dot{\mathcal{V}}}$:
\begin{equation}
    {\footnotesize
    \begin{aligned}    
    \bm{\dot{\mathcal{V}}}\leq&\bm{\sum}_{j=1}^3\!\!\Bigg{\{}\!\!-\!\frac{1}{2}\lambda_{\min}(\bm{Q}_j)\|\bm{\mathcal{E}}_{\bm{\textit{x}} j}\|^2+\lambda_{\max}(\bm{P}_j)(\varepsilon_{\bm{\textit{x}} j}+\frac{\varepsilon_f}{m})\|\bm{\mathcal{E}}_{\bm{\textit{x}} j}\|\\[-5pt]
    &+\frac{\sqrt{3}\,\overset{\tiny \text{max}}{m}\lambda_{\max}(\bm{P}_j)\left(\varepsilon_{u}+\varepsilon_{c}\right)}{m}\|\bm{\mathcal{E}}_{\bm{\textit{x}} j}\|\|\bm{e_R}\|\Bigg{\}}\\
     &\!-\!k_{R}c_{R}\|\bm{e}_{\bm{R}}\|^{2}\!-\!\left(k_{\Omega}\!-\!c_R\right)\|\bm{e}_{\bm{\Omega}}\|^{2}\!+\!k_{\Omega}c_{R}\|\bm{e}_{\bm{\Omega}}\|\|\bm{e}_{\bm{R}}\|\!+\!c_R\|\bm{e}_{\bm{\Omega}}\|^2\\
    &+c_R\bigg{(}\varepsilon_{\bm{\textit{R}}}+\frac{\varepsilon_{\mathbf{M}}}{\lambda_{\min}(\bm{J})}\bigg{)}\|\bm{e}_{\bm{R}}\|+\bigg{(}\varepsilon_{\bm{\textit{R}}}+\frac{\varepsilon_{\mathbf{M}}}{\lambda_{\min}(\bm{J})}\bigg{)}\|\bm{e}_{\bm{\Omega}}\|,
    \end{aligned}
    } 
    \label{dot_V_bound1}
\end{equation}
where $\|\bm{B}\|$ vanishes since $\|\bm{B}\|=1$. The $\varepsilon_{\bm{\textit{x}} j}\!\in\!\mathbb{R}$ is defined as the upper bound of $j^{th}$ optimal approximation error component $\bm{\varpi}^{[j]}_{\textit{x}}$:
\begin{equation}
{\small
\begin{aligned}
\|\bm{\varpi}^{[j]}_{\textit{x}}\|
\leq\varepsilon_{\bm{\textit{x}} j}.
\end{aligned}
}
\label{varpi_bound}
\end{equation}
Applying Young's inequality to the second term of Eq.~\eqref{dot_V_bound1}:
\begin{equation}
    {\footnotesize
    \begin{aligned}    
&\lambda_{\max}(\bm{P}_j)(\varepsilon_{\bm{\textit{x}} j}+\frac{\varepsilon_f}{m})\|\bm{\mathcal{E}}_{\bm{\textit{x}} j}\|\\
&\leq\frac{\lambda_{\max}(\bm{P}_j)^2(\varepsilon_{\bm{\textit{x}} j}+\frac{\varepsilon_f}{m})^2}{\lambda_{\min}(\bm{Q}_j)}+\frac{\lambda_{\min}(\bm{Q}_j)}{4}\|\bm{\mathcal{E}}_{\bm{\textit{x}} j}\|^{2}.\\
    \end{aligned}
    } 
    \label{dot_V_bound2}
\end{equation}
From here, the Eq.~\eqref{dot_V_bound1} can be reformulated as follows:
\begin{equation}
    \begin{aligned}    
    \bm{\dot{\mathcal{V}}}\leq&-\bm{z}^{\top}\!\bm{\mathcal{M}}\,\bm{z}+\mathbf{C},
    \end{aligned}
    \label{dot_V_quadratic}
\end{equation}
where $\bm{z}=\left(\|\bm{\mathcal{E}}_{\bm{\textit{x}} 1}\|,\|\bm{\mathcal{E}}_{\bm{\textit{x}} 2}\|,\|\bm{\mathcal{E}}_{\bm{\textit{x}} 3}\|,\|\bm{e}_{\bm{R}}\|,\|\bm{e}_{\bm{\Omega}}\|\right)^{\top}\!\!\!\in\mathbb{R}^{5}$. The $\mathbf{C}\in\mathbb{R}$ is a constant term given by:
\begin{equation}
    \begin{aligned}    
\mathbf{C}=\mathbf{C}_{\bm{\textit{R}}}+\frac{\lambda_{\max}(\bm{P}_j)^2(\varepsilon_{\bm{\textit{x}} j}+\frac{\varepsilon_f}{m})^2}{\lambda_{\min}(\bm{Q}_j)},
    \end{aligned}
    \label{C}
\end{equation}
where $\mathbf{C}_{\bm{\textit{R}}}$ was introduced in Eq.~\eqref{C_R}.  
The matrix $\bm{\mathcal{M}}\in\mathbb{R}^{5\times5}$ is given by:
\begin{equation}
    \begin{aligned}    
       \bm{\mathcal{M}}={\renewcommand{\arraycolsep}{2pt}\begin{bmatrix}
\bm{\mathcal{M}}_{\bm{\textit{x}}}&\frac{1}{2}\bm{\mathcal{M}}_{\bm{\textit{x}}\bm{\textit{R}}}\\
\\[-2pt]
     \frac{1}{2} \bm{\mathcal{M}}_{\bm{\textit{x}}\bm{\textit{R}}}^{\top} & \bm{\mathcal{M}}_{\bm{\textit{R}}}\\
    \end{bmatrix}},
    \end{aligned}
\end{equation}
where submatrix $\bm{\mathcal{M}}_{\bm{\textit{R}}}$ has been given in Eq.~\eqref{Matrix_R} and 
\begin{equation}
    \begin{aligned}    
       \bm{\mathcal{M}}_{\bm{\textit{x}}}={\renewcommand{\arraycolsep}{2pt}\begin{bmatrix}
\frac{\lambda_{\min}(\bm{Q}_1)}{4}&0&0\\
       0 &\frac{\lambda_{\min}(\bm{Q}_2)}{4}&0\\
       0 &0&\frac{\lambda_{\min}(\bm{Q}_3)}{4}\\
    \end{bmatrix}},
    \end{aligned}
    \label{Mx}
\end{equation}
\begin{equation}
    \begin{aligned}    
      \bm{\mathcal{M}}_{\bm{\textit{x}}\bm{\textit{R}}}={\renewcommand{\arraycolsep}{10pt}\begin{bmatrix}
\frac{-\sqrt{3}\,\overset{\tiny \text{max}}{m}\lambda_{\max}(\bm{P}_1)\left(\varepsilon_{u}+\varepsilon_{c}\right)}{m}&0\\
      \frac{ -\sqrt{3}\,\overset{\tiny \text{max}}{m}\lambda_{\max}(\bm{P}_2)\left(\varepsilon_{u}+\varepsilon_{c}\right)}{m} &0\\
       \frac{-\sqrt{3}\,\overset{\tiny \text{max}}{m}\lambda_{\max}(\bm{P}_3)\left(\varepsilon_{u}+\varepsilon_{c}\right)}{m} &0\\
    \end{bmatrix}}.
    \end{aligned}
    \label{MxR}
\end{equation}
Then, we give the Schur complement of $\bm{\mathcal{M}}_{\bm{\textit{x}}}$ in $ \bm{\mathcal{M}}$: 
\begin{equation}
{\small
    \begin{aligned}    
\bm{\mathcal{M}}/\bm{\mathcal{M}}_{\bm{\textit{x}}}&=\bm{\mathcal{M}}_{\bm{\textit{R}}}-\frac{1}{4}\bm{\mathcal{M}}_{\bm{\textit{x}}\bm{\textit{R}}}^{\top}\bm{\mathcal{M}}_{\bm{\textit{x}}}^{-1}\bm{\mathcal{M}}_{\bm{\textit{x}}\bm{\textit{R}}},\\
&={\renewcommand{\arraycolsep}{2pt}\renewcommand{\arraystretch}{1.5}\begin{bmatrix}
\frac{k_Rc_R}{2}-\Xi&\frac{-k_{\Omega}c_R}{2}\\
       \frac{-k_{\Omega}c_R}{2} &\frac{k_{\Omega}\!-\!c_R}{2}\\
    \end{bmatrix}},
    \end{aligned}
    }
    \label{Schur Complement1}
\end{equation}
where positive term $\Xi\in\mathbb{R_+}$ denotes the stability loss caused by coupling between attitude control on $\mathbf{SO}(3)$ and position control on $\mathbb{R}^3$. Its value is given by:
\begin{equation}
{\small
    \begin{aligned}    
\Xi\equiv3\displaystyle\bm{\sum}_{j=1}^3\frac{\overset{\tiny \text{max}}{m}^2\lambda_{\max}(\bm{P}_j)^2\left(\varepsilon_{u}+\varepsilon_{c}\right)^2}{\lambda_{\min}(\bm{Q}_j)m^2}
    \end{aligned},
    }
    \label{Xi}
\end{equation}
where the magnitude of  $\left(\varepsilon_{u}+\varepsilon_{c}\right)^2$ depends on the  position tracking demand and the disturbance force acting on the quadrotor. In addition, when the physical performance of the quadrotor improves, $\lambda_{\max}(\bm{P}_j)^2/\lambda_{\min}(\bm{Q}_j)$ can be chosen smaller. Therefore, Eq.~\eqref{Xi} essentially describes how the stability loss of the quadrotor is influenced by external disturbance force, position tracking demand, and physical performance.
By appropriately choosing $\bm{Q}_j$, $\bm{\Lambda}_{\textit{x}j}$, $k_R$ and $k_\Omega$, we can make $\Xi$ sufficiently small such that:
\begin{equation}
{\small
    \begin{aligned}
    \Xi\!< \min\left\{\frac{k_Rc_R}{2},\frac{k_Rc_R(k_\Omega-c_R)-k_\Omega^2c_R^2}{2(k_\Omega-c_R)}\right\},\label{Xi bound}
\end{aligned}
}
\end{equation}
where the constant $c_R$ satisfies Eq.~\eqref{cR bound}. Then, the matrix $\bm{\mathcal{M}}$ becomes positive definite since:
\begin{equation}
{\small
    \begin{aligned}    
\bm{\mathcal{M}}_{\bm{\textit{x}}}\succ0,\,\,\text{and}\,\,\bm{\mathcal{M}}/\bm{\mathcal{M}}_{\bm{\textit{x}}}\succ0.
    \end{aligned}
    }
    \label{Positive definite condition of M}
\end{equation}
Therefore, Eq.~\eqref{dot_V_quadratic} can be further expressed as:
\begin{equation}
    \begin{aligned}    
    \bm{\dot{\mathcal{V}}}\leq-\lambda_{\min}(\bm{\mathcal{M}})\|\bm{z}\|^2+\mathbf{C},
    \end{aligned}
    \label{dot_V_R_quadratic2}
\end{equation}
where $\mathbf{C}>0$. To proceed, substituting Eq.~\eqref{V_p1p2bound}, it holds that:
\begin{equation}
    \begin{aligned}    
    \bm{\dot{\mathcal{V}}}\leq-2\beta_2\bm{\mathcal{V}}+\mathbf{C},
    \end{aligned}
    \label{dot_V_quadratic2}
\end{equation}
with $\beta_2=\frac{\lambda_{\min}(\bm{\mathcal{M}})}{2\mathfrak{p}_4\lambda_{\max}(\bm{\mathfrak{M}}_{2})}$. This drives that:
\begin{equation}
    \begin{aligned}    
    \|\bm{z}(t)\|\leq\alpha_2\|\bm{z}(0)\|e^{-\beta_2 t}+\epsilon_2,
    \end{aligned}
    \label{NES condition1}
\end{equation}
where $\bm{z}(0)\!\in\!\mathcal{D}_0$, $\alpha_2=\sqrt{\frac{\mathfrak{p}_4\lambda_{\max}(\bm{\mathfrak{M}}_{2})}{\mathfrak{p}_3\lambda_{\min}(\bm{\mathfrak{M}}_{1})}}$, $\epsilon_2=\sqrt{\frac{\mathbf{C}}{2\beta_2\mathfrak{p}_3\lambda_{\min}(\bm{\mathfrak{M}}_{1})}}$. 

Similar to the \textbf{\textit{Proof of Proposition 1}}, the universal approximation theorem \cite{1989 Multilayer feedforward networks are universal approximators} ensures arbitrarily small upper bounds of the approximation errors, i.e., $\varepsilon_{\bm{\textit{x}}j}\to0^+$. Furthermore, precise identification of the reference coefficient $c'_T$ through aerodynamic experiments ensures that the upper bound $\varepsilon_f$ reaches zero. With the fact that $\mathbf{C}_{\bm{\textit{R}}}\to0^+$, $\varepsilon_{\bm{\textit{x}}j}\to0^+$,  $\varepsilon_{f}\to0$, it holds that $\mathbf{C}\to0^+$ and $\epsilon_2\to0^+$. This guarantees that the equilibrium of all state errors $\{\bm{\mathcal{E}}_{\bm{\textit{x}} j}\}_{1\leq j \leq3}, \bm{e}_{\bm{R}},\bm{e}_{\bm{\Omega}}$ is near-exponentially stable. Since $\{\bm{\mathcal{E}}_{\bm{\textit{x}} j}\}_{1\leq j \leq3}$ are the components of $\bm{e}_{\bm{x}},\bm{e}_{\bm{v}}$, it holds that \textbf{\textit{Proposition 2}} is established.
In addition, since all state errors are bounded within the attraction domain $\mathcal{D}_0$, there exists a smaller compact set such that $\bm{\mathcal{C}}\subset\mathcal{D}_0$ to establish \textbf{\textit{Proposition 0}}.

\vfill
\end{document}